\documentclass{aa}  
\usepackage{graphicx}
\usepackage{txfonts}
\usepackage{xcolor}
\usepackage{soul}
\usepackage{mathtools}
\usepackage[normalem]{ulem}
\begin{document}

   \title{Particle acceleration at recollimation shocks in subrelativistic jets}

   \subtitle{A model for jets in Seyfert galaxies, microquasars, and protostellar systems}

   \author{E. Peretti
          \inst{1}\inst{2}\fnmsep\thanks{enrico.peretti.science@gmail.com}
          \and 
          E. Amato \inst{1}
          \and
          S.~S.~Cerri \inst{3} 
          \and
          G. Morlino \inst{1}
          \and 
          L.P. Pullano \inst{1}\inst{4}
          \and 
          S. Recchia \inst{1}\inst{5}
          }

   \institute{INAF, Istituto Nazionale di Astrofisica, Osservatorio Astronomico di Arcetri, Largo E. Fermi 5, 50125 Florence, Italy 
         \and
    Astroparticule et Cosmologie, Universite Paris Cit\'e, 10 Rue Alice Domon et L\'eonie Duquet, F-75013 Paris, France
        \and 
    Université Côte d’Azur, Observatoire de la Côte d’Azur, CNRS, Laboratoire Lagrange,
    Bd de l’Observatoire, CS 34229, 06304 Nice cedex 4, France
    \and 
    University of Pisa, Lungarno Antonio Pacinotti, 43, 56126 Pisa, Italy
    \and
    Institute of Nuclear Physics Polish Academy of Sciences, PL-31342 IFJ-PAN, Krakow, Poland
             }

   \date{Received XXX, XXX; accepted XXX, XXX}

  \abstract
    {Growing observational evidence suggests that subrelativistic astrophysical jets may accelerate particles at slowly evolving standing shocks. Recollimation shocks are expected to develop when jets expand in dense environments; their formation may be mediated by the pressure of the cocoon surrounding the jet, while remaining compatible with a quasi-stationary behavior. Despite their high inclination relative to the jet axis, such shocks can be strong and can enable efficient particle acceleration.}
   {The aim of this work is to improve the general understanding of particle acceleration via diffusive shock acceleration at recollimation shocks by developing a versatile modeling framework applicable to different classes of astrophysical jets, including Seyfert galaxies, microquasars, and protostellar systems.}
   {We extended an analytic jet hydrodynamics model previously introduced in the literature to the subrelativistic regime and used it to identify the expected locations of the recollimation shock and the jet head. Within this framework, we formulated a semi-analytic acceleration and transport model for particles injected at the recollimation shock via diffusive shock acceleration.}
   {By solving the space-dependent transport equation, we obtained particle distributions and spectra along the jet, as well as robust predictions for the maximum energies achievable as a function of the intrinsic properties of the system and the source class.}
   {Our results indicate that recollimation shocks may play a central role in particle acceleration in subrelativistic jets. In Seyfert galaxies such shocks may accelerate particles from PeV up to EeV energies, while in microquasars and protostellar jets maximum energies of respectively tens of PeV and up to TeV energies are expected. While leptonic emission may be associated with bright knots along the jet, accelerated protons are expected to escape the jet and interact with the surrounding cocoon. Depending on the properties of the system, proton interactions can give rise to extended hadronic emission morphologies.}


   \maketitle
%
%-------------------------------------------------------------------

\section{Introduction}

Astrophysical systems can produce collimated outflows, or jets, across a wide range of power and spatial scales.
The acceleration and collimation of these jets can result from different physical mechanisms, including geometric collimation by dense ambient material, magnetic tension, recollimation shocks of different nature, or magnetorotational effects.
The extraction of rotational energy from compact objects \citep{Penrose1969,Blandford-Znajek1977} and magnetocentrifugal disk winds \citep{Blandford-Payne1982} have long been considered among the most efficient launching mechanisms of jets. 
Early magnetohydrodynamic (MHD) simulations demonstrated the feasibility of jet formation in magnetized disks \citep{Shibata_1986}, and more recent general relativistic MHD studies have shown that, in magnetically arrested disks, the outflow luminosity can exceed the accretion power in agreement with the energy extraction resulting from the Blandford–Znajek process \citep{Tchekhovskoy2011}.

As soon as jets were understood to be highly supersonic, the presence of strong shocks was promptly postulated, and consequently diffusive shock acceleration (DSA) was soon considered as a viable mechanism to accelerate particles in these objects.
Pioneering theoretical works on particle acceleration and radio emission established the foundations for interpreting jet phenomenology \citep{Blandford-Rees1974,Scheuer1974,Blandford-Rees1978,Bladford_Konig1979}. 
The first hydrodynamic simulations then clarified the basic structure of supersonic jets and characterized the shocked environments, as well as the formation of wide regions---typically referred to as cocoons---surrounding the jets themselves \citep{Norman1982}. 
The propagation of shocks in jets has been proposed as a key ingredient in explaining spectral evolution in jets of active galactic nuclei (AGN), such as blazars \citep{Marscher_1985}, while subsequent relativistic simulations examined the role of external pressure gradients \citep{Daly_Marscher1988}, jet confinement induced by the cocoon \citep{Begelman1989}, and the formation of standing shocks \citep{Gomez1995,Gomez2000}. 
Relativistic effects such as superluminal motion \citep{Gomez1997} and the emergence of shock-induced knots \citep{Kommissarov1997} were naturally reproduced. Analytical treatments of jet expansion  in active galaxies \citep{Kaiser-1997} and time-dependent MHD simulations \citep{Ouyed1999} further constrained jet stability, while nonrelativistic shocks triggered by Kelvin–Helmholtz instabilities were shown to efficiently accelerate electrons \citep{Micono1999}. 

Important progress on  jet dynamics has also been achieved  through analytic studies. 
In particular, analytic models of jet–cocoon systems indicate that the degree of collimation depends sensibly on the ambient conditions, predicting either tightly collimated or freely expanding outflows \citep{Krause_2003,Bromberg_2011}. 
This behavior was found to be consistent with 2D relativistic MHD simulations, which found that axial magnetic fields strengthen recollimation shocks, whereas predominantly toroidal fields tend to weaken them \citep{Mizuno_2015}. 
Three-dimensional MHD simulations have shown that kink instabilities can generate shocks and produce the knotty structures observed in many jets \citep{Moll_2008}. 
In addition, when radiative losses were included, simulations also showed pronounced flow focusing downstream of the recollimation region \citep{Bodo2018}. Interestingly, interactions between jets and embedded obstacles were suggested to locally boost nonthermal emission \citep{Bosch-Ramon_2015_obstacles}.

Collimated jets in astrophysical systems have been observed over a wide range of speeds, from $10^2 \, \rm km \, s^{-1}$ in the case of protostars,  to Lorentz factors $\Gamma \gg 1$ in the case of AGN and gamma-ray bursts. 
However, relativistic shocks are known to be inefficient in accelerating particles, and hence here we focus on shocks from nonrelativistic jets, and we limit our discussion to those sources featuring bulk Lorentz factors $\Gamma \lesssim 2$.
These are mainly found in the objects presented below.
{Mildly relativistic shocks are extensively studied through kinetic plasma simulations, including particle-in-cell and hybrid methods, which are essential in order to capture the microphysics of particle acceleration and magnetic turbulence generation in collisionless plasmas~\citep[e.g.,][]{Sironi2011,Sironi_2011_2,Caprioli2014,Jikey2026,Minh2026}.}
{Finally, we note that, even in relativistic flows, shock obliquity can lead to a subrelativistic velocity of the upstream plasma in the shock frame, although such configurations may correspond to superluminal shocks~\citep{Sciaccaluga2025}.}

{Microquasars:}
Powered by the accretion of a stellar-mass compact object from a companion star, microquasars are the prototypical Galactic objects where jets are observed with both mildly relativistic and highly relativistic velocities.
Apparent superluminal motions were first identified in these systems by \citet{Mirabel1994,Mirabel1998}. Their nonthermal emission has been interpreted through various leptonic \citep{Band_1986,Atoyan_1999,Romero_2002} and hadronic scenarios \citep{Romero2003}. Several models address cosmic-ray (CR) production in microquasar environments \citep{Heinz_2002}. Multi-zone leptonic and lepto-hadronic frameworks have been applied to a range of sources \citep{Bosch-Ramon_2004,Bosh-Ramon_2004b,Markoff_2005,Bosch-Ramon_2005,Romero_2008MQp,Vila_2010_MQs,Sudoh_2020,Kimura_2020,Kantzas2021,Khangulian_2024,Wan_2025}. 
Hydrodynamic simulations have examined shock formation (bow, termination, and recollimation) \citep{Bordas_2009} and jet propagation in complex environments \citep{Bosh-Ramon_2011}. 
High-energy variability, absorption, and pair production have been extensively studied, particularly in Cygnus X-3 \citep{Dubus_2010,Zdziarski_2012_cygnusX3,Cerutti2011_cygnusx3,Tavani_2009_CygnusX3,Fermi_LAT_2009_CygnusX3}. 
In these systems, stellar winds may trigger jet launching even in the absence of a stable accretion disk \citep{Barkov_2012}. 

SS433 is another Galactic microquasar featuring extremely interesting and complex features. 
In addition to the mildly relativistic jet resulting from a super-Eddington accreting stellar-mass compact object, it is characterized by a wide opening angle mildly relativistic outflow typical of ultra-luminous X-ray sources \citep{Middleton_SS433}. 
Such observations led to recent interesting hypotheses on the nature of the surrounding nebula, W50, as the result of superposition of a disk wind interacting with the jet \citep{Churazov_2024}. This interpretation seems to be supported by diffuse X-ray emission observed by e-Rosita \citep{Sunyaev_2025}.
However, the nature of the SS433/W50 nebula is still highly uncertain. 
Intermittent jet activity has also been suggested for SS433 \citep{Goodall_2011}, whose morphology has also been reproduced through precessing-jet simulations \citep{Meliani2014_SS433} and analytic descriptions of recollimation shocks \citep{Yoon2016}. 
MHD simulations of the SS433/W50 system have also reproduced the observed nebular structure \citep{Ohmura_2021}.

Interestingly, microquasars have  recently been detected at very high energies (100 GeV - 100 TeV) by the High Altitude Water Cherenkov Observatory (HAWC) and the High Energy Stereoscopic System (HESS) \citep{HAWC_2018_SS433,HESS_SS433_2024,HESS_V4641_2025}, and by the Large High Altitude Air Shower Observatory (LHAASO) up to hundreds of TeV \citep{LHAASO_BH}, thereby suggesting that they are Galactic PeV accelerators.
In addition, X-ray observations of SS433 provided further support to the presence of a strong recollimation shock \citep{Tsuji_2025}.
These objects were already believed to contribute to Galactic CRs through standard proton \citep{Romero2003} or neutron escape \citep{Escobar_2021,Escobar_2022}. 
However, the above-mentioned recent observational evidence in the TeV domain stimulated novel investigations.
Microquasar jets \citep{Abaroa_2024,Bykov_2025} and super-Eddington outflows have been proposed as potential Galactic PeVatrons \citep{Peretti_2025_ULX}. 
The contribution of microquasars to the Galactic CR population has been re-examined in several recent works \citep{Peretti_2025_ULX,Wang_2025,Zhang_2025,Kaci2025}. Finally, the remnants of microquasars have also been proposed as   some of the extended unidentified LHAASO sources \citep{Abaroa26_MQR}.

{Active galaxies:} 
{The accretion of matter onto supermassive black holes in the core of active galaxies can produce feedback of a different nature. 
The most powerful, and perhaps most iconic, are the highly relativistic (with bulk Lorentz factors typically ranging from a few up to a few tens) and collimated jets observed in radio galaxies and blazars \citep{Urry-Padovani_1995}, and over the last two decades also in narrow-line Seyfert 1 galaxies \citep{Mohave_2016,Foschini_2020,Foschini_2026}. 
The presence of such powerful relativistic jets in active galaxies is somewhat rare compared to the whole AGN population, in that the large majority ($\sim$ 90\%) of AGN are typically nonjetted,  either without clear indications of relativistic outflows or featuring a relatively weak or subrelativistic jet \citep{Padovani_2017}.
In contrast to these rare and highly relativistic jets, a more ubiquitous form of feedback is represented by slower outflows.} 
Launched from the nearest neighborhoods of supermassive black holes, mildly relativistic and Newtonian outflows are observed in several AGN \citep{Veilleux_05}.
Examples include Seyfert galaxies such as NGC 1068 \citep{Gallimore96_1,Gallimore96_2,Lenain2010,Salvatore_2023_1068}, NGC~4151 \citep{Ulvestad_2005,Willams_2017}, and NGC 6764 \citep{Kharb_2010}, where the jets display velocities in the range $\sim 0.01-0.1c$ and possibly also evidence of precession or kinks.  
Jet-driven acceleration of molecular gas to several hundred km s$^{-1}$ has been reported in IC 5063 \citep{Tadhunter2014}. 
{\citet{Costa2024_FR0} performed 3D hydrodynamic simulations of weak AGN jets---typical of Fanaroff-Riley 0 objects~\citep{Ranieri2023}---and found that after a complex of recollimation and reflected shocks, strong instabilities might develop and lead to a deceleration of the jet.}
Recollimation shocks have also been invoked to explain nonthermal components in narrow-line Seyfert 1 jets \citep{Doi_2018,Hada2018}, while simulations of nonrelativistic AGN jets have explored cocoon formation \citep{Ohmura2020}. 
Interestingly, particle acceleration may also take place in backflows surrounding relativistic jets and achieve the highest energies in the cosmic-ray spectrum \citep{Matthews_2019}.
In addition to {relatively weak collimated jets,} nonjetted AGN can power wide opening angle winds reaching mildly relativistic velocities where particle acceleration up to the ultra-high-energy domain is thought to take place \citep{Wang_2017,Peretti_2023,Ehlert_2025,Peretti_2025_NGC4151,Baptiste2026}.

{Protostellar jets:} 
The possibility that young stellar objects, or protostars, could launch fast and collimated jets followed pioneering works on nebulous objects observed around young stars \citep{Herbig_1951,Haro_1951}. 
Protostellar jets represent another class of nonrelativistic collimated outflows that, different from the previous cases of microquasars and AGN, do not typically exceed velocities of $\sim 10^3 \, \rm km \, s^{-1}$. 
Early works on Herbig–Haro (HH) systems examined their dynamics and the impact of strong radiative cooling \citep{Blondin_1989}. Observations of synchrotron emission in T Tauri jets suggest the presence of ordered magnetic fields that may assist collimation \citep{Ray1997_TTauri}. 
Jets launched by young stellar objects are expected to play an important role in angular momentum regulation \citep{Konigl2000}. Radio surveys have established their extension up to parsec scales \citep{Reipurth_2004}. High-energy emission has been modeled through leptonic and hadronic processes \citep{Araudo_2007}, and numerical simulations have investigated the variability of radiative HH jets \citep{Raga_2007}. Massive protostellar jets may produce gamma rays through particle acceleration at their terminal and reverse shocks \citep{Bosch-Ramon2010_protostar}. 
Evidence for efficient acceleration includes high ionization rates \citep{Padovani2015}, radio signatures compatible with diffusive shock acceleration \citep{Rodriguez-Kamentzky_2016}, and observational hints of recollimation shocks \citep{Rodriguez-Kamentzky2017,Carrasco-Gonzalez_2021}. 
Particle acceleration at protostellar jet heads has been studied in detail \citep{Araudo2021_protostar}, and recent gamma-ray detections of protostellar systems further support this picture \citep{Yan_2022,Yang_2023,Wilhelmi_2023}.\\

Despite great progress in our understanding of jet dynamics and radiation, our understanding of particle acceleration in such systems is still poor.
This gap between jet dynamics and multi-messenger radiation is  the subject of the present paper.
In particular, for this work we developed a model of DSA at recollimation shocks induced by cocoons in the context of subrelativistic jets. 
This is performed in the context of an extension of the analytic hydrodynamic framework developed by \cite{Bromberg_2011}.
The paper is organized as follows. 
Section ~\ref{Sec: dyn-and-structure} is devoted to the jet structure and evolution. In Sect.~\ref{Sec: Transport} we describe the particle acceleration model and in Sect.~\ref{Sec: Results} we present our results focusing, in particular, on the maximum energy for the accelerated particles. 
We discuss the limitations of our model and future prospects in Sect.~\ref{Sec: discussion}. We   draw our conclusions in Sect.~\ref{Sec: Conclusions}.

\section{The jet-cocoon system}
\label{Sec: dyn-and-structure}

\begin{figure}[t]
    \centering   \includegraphics[width=0.90\columnwidth]{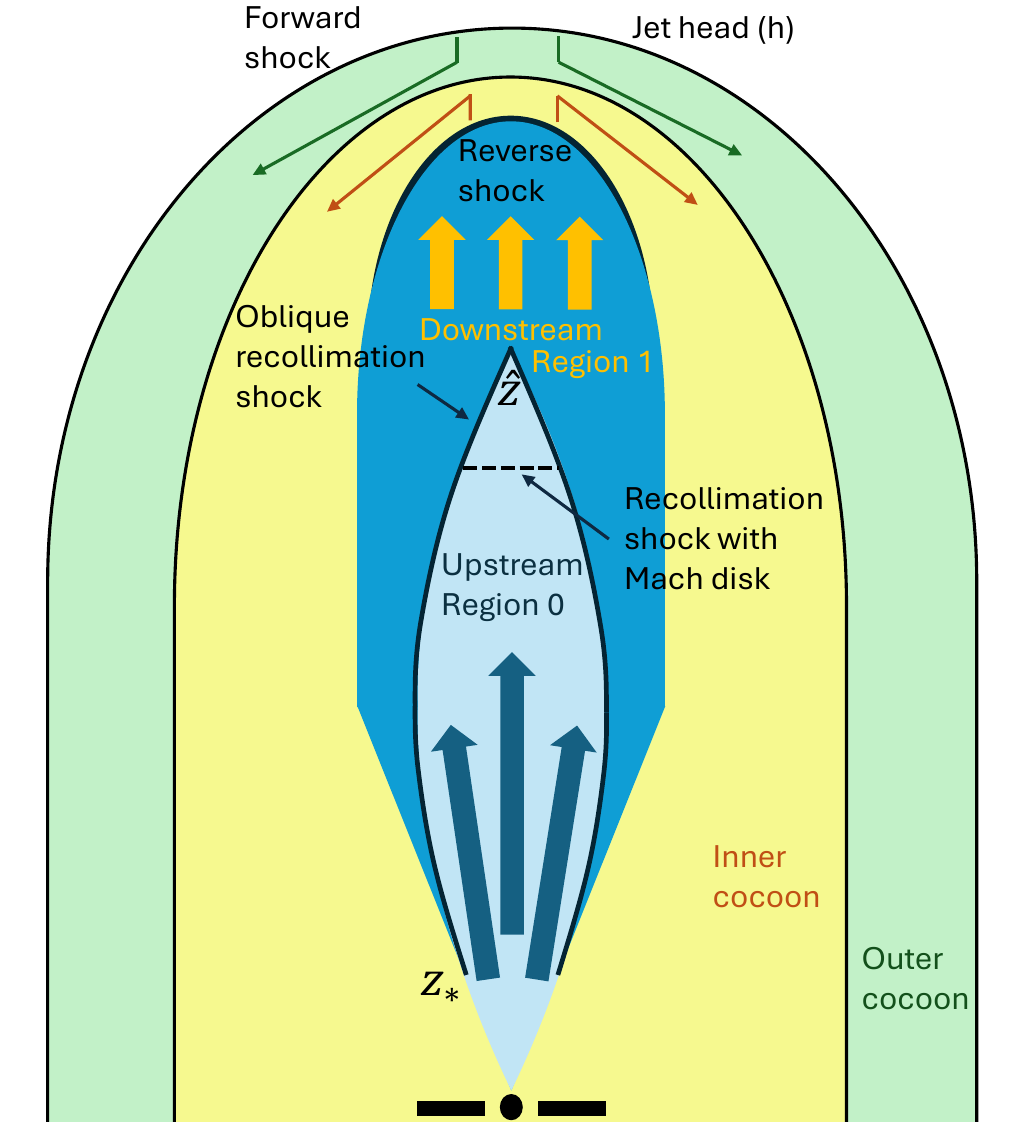}
    \caption{Sketch illustrating the jet-cocoon system. Particles are accelerated at the recollimation shock and then advected in the jet. The jet material that passes through the reverse shock fills the inner cocoon (yellow area), while the external medium reached by the jet head and, laterally, by the cocoon surface, is swept up and accumulated in the outer cocoon (green area). {Two alternative recollimation scenarios are illustrated: (1) a complete cusp and  (2) a Mach disk.}}
    \label{Fig: JET sketch}
\end{figure}

Collimated jets launched by compact objects are characterized by complex dynamics regulated by the interaction with the surrounding medium.
In particular, when a jet is launched with supersonic velocity, it develops a terminal region known as the jet head. 
Such a jet head is composed by a forward shock, which separates the undisturbed external medium from the shocked one, a contact discontinuity, which sets the separation between the shocked ambient medium and the shocked jet material, and a reverse shock, where the high-velocity jet plasma is shocked. 
The shocked jet material and the shocked ambient medium typically flow sideways with respect to the jet axis and fill a hot and high-pressure cocoon. In particular, inner and outer cocoons are inflated respectively by the shocked jet material and the shocked ambient material. 
The cocoon surrounds the jet and can contribute to confine its opening angle beyond a certain distance from the central engine. 
In particular, if the component of the ram pressure perpendicular to the jet axis falls below the thermal pressure in the cocoon, a recollimation shock forms and the jet is forced to change its geometry from conical to cylindrical.  
A sketch of the jet-cocoon system is shown in Fig.~\ref{Fig: JET sketch}, {where $z$ is taken as the spatial coordinate along the jet axis. As illustrated in the sketch, the recollimation can feature different shapes: (1) a smooth cusp or (2) a truncated cusp with a Mach disk. The details of the shock structure are provided in Sect.~\ref{Subsec: recollimation}}.

Newtonian jets from microquasars, AGN and some protostars are expected to be characterized by a plasma density smaller than that of the medium in which they are expanding. 
This results in a dominant role played by the cocoon in the jet collimation, and consequently leads to the formation of a recollimation shock \citep{Bromberg_2011}.
Recollimation shocks are generally characterized by a certain angle between the fluid motion and the normal to the shock. In other words, they are oblique with respect to the flow propagation direction.
Nevertheless, they can be strong enough to efficiently accelerate particles as we show below. 
{In addition, if the jet cannot adjust smoothly to the cocoon overpressure, the recollimation shock may collapse into a sharp Mach disk structure, which is expected to be stronger and more efficient at dissipating energy than a purely oblique shock.}

Although astrophysical jets close to their central engine are expected to be magnetically dominated and with strong helicoidal magnetic fields that help in the collimation, we are interested in the properties of these systems at larger scales.
At large distance from the central engine jets are expected to become kinetically dominated because of the rapid dissipation of Poynting flux
\citep[see, e.g.,][]{Giannos2006}. 
In general, the ram pressure is expected to dominate over the thermal pressure which, in turn, is expected to be larger than the magnetic pressure.
This makes the assumption of a pure hydrodynamic model qualitatively justified.
In this context, we develop an analytic hydrodynamic model for the subrelativistic jet-cocoon system inspired by the formalism of \citet[see also \citealt{Krause_2003}]{Bromberg_2011}.

\subsection{Jump conditions at the recollimation shock}
\label{Subs: Jump conditions}

Recollimation shocks are expected to be elongated, and {unless they feature a Mach disk,} to form cusps along the jet axis.
Consequently, the upstream plasma impacts the shock at a certain angle with respect to the shock-normal.
After crossing the shock, only the velocity component parallel to the shock-normal is reduced, while the tangential component remains unaltered. 
As a consequence, the plasma flow is deflected approximately parallel to the jet axis in the downstream --- hence, the name recollimation shock.
Thus, the Rankine-Hugoniot jump conditions only hold for the normal component of the flow $u$, hereafter denoted by a subscript $n$ {(clearly $u_n = u$ for Mach disks)}. 
This results in the following definition for the compression ratio
\begin{equation}
    \label{eq: compression ratio}
    r = \frac{u_{0,n}}{u_{1,n}} = \frac{(\gamma+1) M_{0,n}^2}{(\gamma-1)M_{0,n}^2 +2} 
,\end{equation}
where the subscript 0 ($1$) indicates that the quantity is computed in the upstream (downstream), $\gamma$ is the adiabatic index of the gas, and $M_{0,n}$ is the upstream Mach number associated with the normal component of the flow.  
Mach numbers of mildly relativistic and subrelativistic jets typically range from a few up to a few tens. 
This is enough to make the shock an efficient accelerator, while, as we discuss in Sect.~\ref{Sec: discussion}, it might be not enough to guarantee spectra as hard as $\sim E^{-2}$. 

\begin{figure}[t]
    \centering
    \includegraphics[width=1.0\columnwidth]{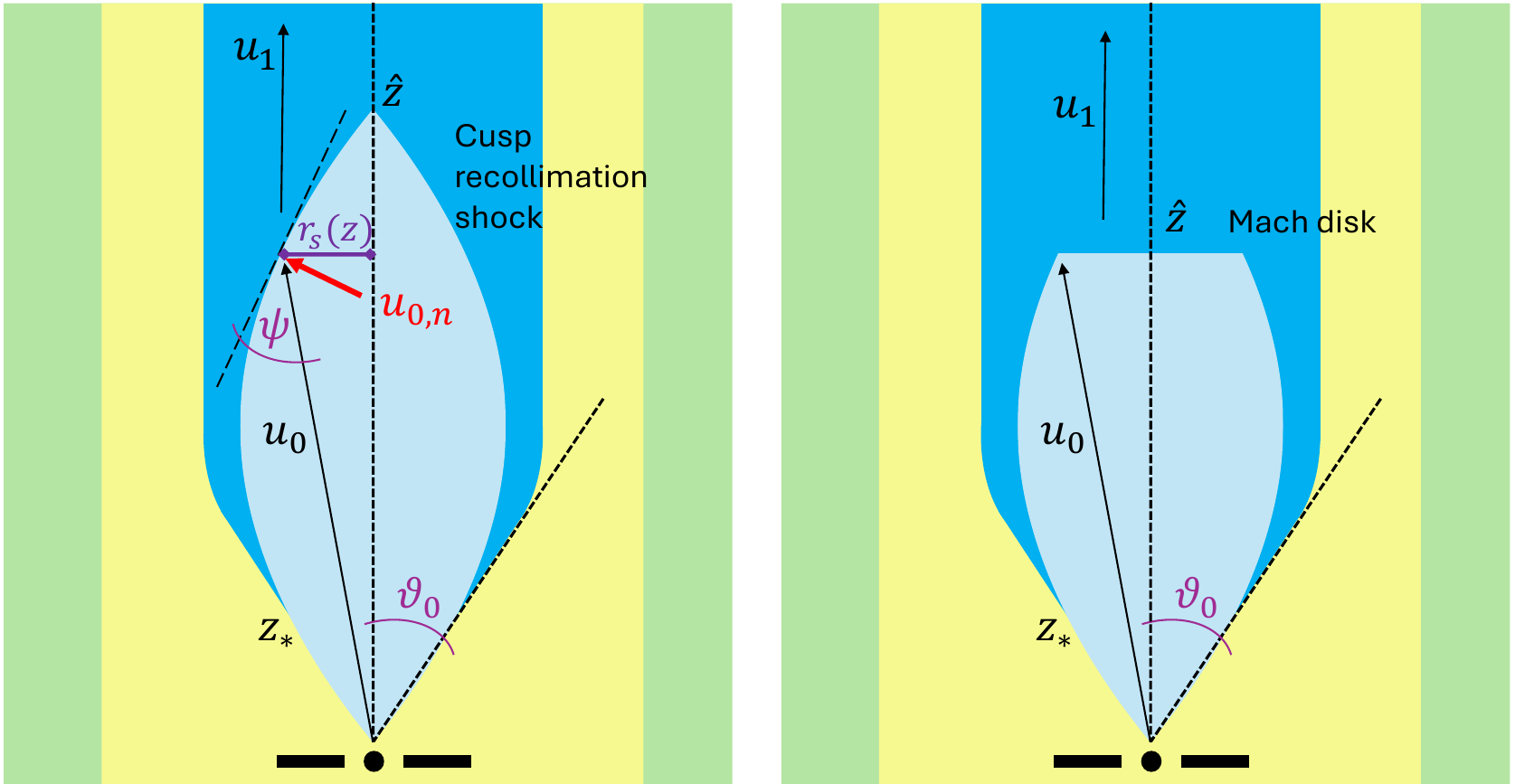}
    \caption{Sketch illustrating the recollimation shock geometry. 
    {\textit{Left}: Cusp-shaped recollimation shock.} The upstream fluid with velocity $u_0$ impacts the shock with angle $\psi$ and is deflected along the jet axis. The component normal to the shock of the upstream velocity is $u_{0, {\rm n}}$. The downstream plasma speed is $u_1$. The shock radius, $r_s$, is the radius of the shock which approximately coincides with the jet radius at $\hat{z}/2$. The recollimation shock forms at $z_*$. {\textit{Right}: Recollimation shock with Mach disk. For a sizable portion of the structure, the shock is orthogonal to the jet axis and, consequently, to the jet flow.}}
    \label{Fig: Recollimation sketch}
\end{figure}

The inclination of the recollimation shock is extremely relevant for the characterization of the accelerator properties as well as the jet structure.
In fact, higher inclinations between the shock normal and the velocity vector result in lower acceleration efficiency for the accelerated particles, because the pressure of accelerated particles is expected to be a fraction of the ram pressure component normal to the shock.
Similarly, also the jet thermal pressure downstream of the shock, $P_1$, strongly depends on the angle. 
From the pressure jump condition one can express the pressure as
\begin{equation}
\label{eq: jet-pressure-postshock}
    P_1 =  \frac{(r-1)}{r} \sin^2{\psi}\, \rho_0 u_{0}^2 = k_p \, \rho_1 u_{1}^2, 
\end{equation}
where $\psi$ is the angle between the shock surface and the upstream velocity vector (see {the left panel of} Fig.~\ref{Fig: Recollimation sketch}), $\rho_{0(1)}$ represents the upstream (downstream) density, and $k_p$ is the  thermal-to-ram pressure ratio in the downstream plasma, which can be expressed as
\begin{equation}
    k_p = \frac{r-1}{1+r^2 \cot^2\psi} \,.
\end{equation}

We parameterize with $k_p$ the impact of the shock inclination on the downstream pressure and recall that typically it can range from 3 in the case of strong shock and large inclination angle (i.e., $\psi \approx 90^{\circ}$), down to $k_p \ll 1$ for $\psi \ll 90^{\circ}$.
{As illustrated in the right panel of Fig.~\ref{Fig: Recollimation sketch}, a Mach disk structure presents the most ideal condition in terms of dissipation, with a locally maximized $k_p$. Such a shock does not necessarily extend for the whole jet section. Therefore, even in presence of a Mach disk, one can expect $k_p \lesssim O(1)$.}
It is worth noticing that the strength of the reverse shock at the jet head is linked to the recollimation shock inclination. Using Eq.~\eqref{eq: jet-pressure-postshock} we can express the Mach number at the reverse shock as
\begin{equation}
    M_1 = \frac{u_1}{c_{s,1}} = \frac{1}{\sqrt{\gamma k_p}}
\end{equation}
such that a small $\psi$ value results in a strong reverse shock, and {vice versa}.
In general, we expect $k_p \ll 1$ and the reverse shock to be strong.

\subsection{Jet characterization}

The motion of the jet head is regulated by the pressure balance between the jet (ram pressure and thermal pressure) and the ram pressure of the external medium
\begin{equation}
\label{eq: head motion 1}
    \rho_{1} (u_{1}-v_h)^2 + P_{1} = \rho_{\rm A} v^2_{h},
\end{equation}
where $\rho_A=n_Am_p$ is the density of the external medium and $v_h$ is the velocity of the jet head.
In this equation we neglected the role played by the thermal pressure of the external medium on the right hand side.
Adopting the same formalism as \citet{Bromberg_2011}, we define $\tilde{L}$ as
\begin{equation}
\label{Eq: L-tilde definition}
    \tilde{L}(t) = \frac{\rho_1}{\rho_{\rm A}} = \frac{2 L_j}{ \rho_{\rm A} u_1^3 \Sigma_1(t) \zeta},
\end{equation}
where $L_j$ is the jet kinetic luminosity, $\Sigma_1$ is the jet section after the recollimation shock, and $\zeta$ is a parameter of order unity that encompasses the normalized contribution of ram pressure and thermal pressure to the total jet kinetic power,
\begin{equation}
    L_j = \Sigma_0 \frac{1}{2} \rho_0 u_0^3 = \Sigma_1 \left[ \frac{1}{2} \rho_1 u_1^3 + \frac{\gamma}{(\gamma-1)} P_1 u_1 \right] = \Sigma_1 \frac{1}{2} \rho_1 u_1^3 \zeta, 
\end{equation}
so that $\zeta = [1+2 \gamma k_p/(\gamma -1 )]$. For $\gamma = 5/3$ depending on the recollimation shock strength it is possible to see that $1 \leq \zeta \leq r^2$, and $\zeta \approx 1$ when $k_p \ll 1$.

By solving Eq.~\eqref{eq: head motion 1} with the use of Eq.~\eqref{eq: jet-pressure-postshock}, we find the following expression for the jet head speed:
\begin{equation}
\label{eq:v_h_full}
    v_h(t) = u_1 \frac{(1+k_p)}{1 + \sqrt{1+(1+k_p)(\tilde{L}^{-1}(t)-1)}}.
\end{equation}
We note that for $k_p \ll 1$ we recover the  solution from \citet{Bromberg_2011} for a subrelativistic jet, i.e., $v_h(t) \approx u_1/(1+\tilde{L}^{-1/2})$. In such a limit, the expression for $v_h$ is valid for all values of $\tilde{L}$, while for a nonnegligible $k_p$ the existence of a physical solution requires $\tilde{L}\leq (1+k_p)/k_p$.  

In the limit $\tilde{L} \ll 1$ the head velocity reads
\begin{equation}
\label{eq: jet-head speed}
    v_h(t) \sim u_1 \tilde{L}^{1/2}(t) \sqrt{1+k_p}.
\end{equation}

\subsubsection{Cocoon}

If strong shock conditions are realized at the reverse shock, the hot shocked gas flows orthogonally to the jet axis and leaves the jet head feeding the inner cocoon, namely the region of shocked jet material that surrounds the whole jet structure from the head to the central engine. 
The inner cocoon is then surrounded by the outer cocoon, namely the ambient gas that has been swept-up and shocked by the forward shock at the jet head, or possibly also by the lateral expansion of the cocoon itself.

The energy injected by the jet in the cocoon can be parameterized as 
\begin{equation}
\label{eq: Cocoon Energy}
    E_{\rm c} \approx \eta L_{j} t,
\end{equation}
where $\eta<1$ is an efficiency parameter accounting for the conversion of jet kinetic energy into pressure at the reverse shock and the amount of energy, $E_j$, still stored in the jet. 
While such energy can be relevant at early times, when the cocoon is yet to be formed, at later times, it becomes subdominant because $E_j/E_c = (L_j z_h/u_1)/(\eta L_j t) \approx v_h/(\eta u_1)$ which is $\ll 1$ if $\tilde{L} \ll 1$ (see Eq.~\ref{eq: jet-head speed}).

The vertical expansion of the cocoon is set by the motion of the jet head along the jet axis, while the lateral expansion velocity, $v_c$, is set by the pressure balance between the external ram pressure and the pressure in the cocoon, so that one has $v_c(t) = \sqrt{P_c(t)/\rho_{\rm A}}$.
By approximating the cocoon as a cylinder, it is possible to estimate its volume as 
\begin{equation}
\label{eq: cocoon volume}
    V_c(t) = z_h(t) \pi R_c^2(t) = \pi C B^2 t^{\alpha+2\beta} = \pi v_h(t) v_c^2(t) t^3 /\alpha \beta^2 ,
\end{equation}
where we assumed a self-similar scaling for the cocoon height --- $z_h(t) = \int dt \, v_h(t) = C t^{\alpha}$ --- and radius --- $R_c(t)=\int dt \, v_c(t) = B t^{\beta} $. $C$ and $B$ are constants to be determined, while $\alpha$ and $\beta$ are self-similar indexes.
By using Eqs.~\eqref{eq: Cocoon Energy}, \eqref{eq: cocoon volume}, and \eqref{eq:v_h_full}, the pressure in the cocoon can be expressed as 
\begin{align}
    \label{eq: cocoon pressure}
    P_c (t) &= (\gamma-1) \frac{E_c(t)}{V_c(t)} = \\ 
    &= \frac{\left[1+\sqrt{1+(1+k_p)(\tilde{L}^{-1}(t)-1)}\right]^{1/2}}{t} \left[ \frac{(\gamma -1) \eta L_j \rho_{\rm A} \alpha \beta^2}{\pi u_1 (1+k_p)} \right]^{1/2} \nonumber \\
    & \xrightarrow[\tilde{L} \ll 1]{}
    \frac{\tilde{L}^{-1/4}(t)}{t} \left[ \frac{ L_j \rho_{\rm A}}{3 \pi u_1} \right]^{1/2} 
    \left[ \frac{3 (\gamma -1) \eta \alpha \beta^2}{(1+k_p)^{1/2}} \right]^{1/2},
\end{align}
where the latter expression is obtained for $\tilde{L} \ll 1$.
We note that our result is equivalent to Eq.~(7) from \citet{Bromberg_2011}. This is especially clear in the limit $\tilde{L} \ll 1$.

\subsubsection{Recollimation shock}
\label{Subsec: recollimation}

The recollimation shock is located where the upstream ram pressure normal to the shock surface matches the pressure in the cocoon:
\begin{equation}
\label{Eq: Pressure-balance recollimation}
    \rho_0 u_{0,n}^2=\rho_0 u_0^2 \sin^2{\psi} \approx P_c.
\end{equation}
As the upstream plasma is freely expanding with a certain opening angle $\theta_0$, the ram pressure, $\rho_0 u_0^2$, is expected to drop approximately as $z^{-2}$.
Following \citet{Bromberg_2011}, for small angles $\psi$, the following relation holds:
\begin{equation}
\label{Eq: proxi angle}
    \sin{\psi} \approx \frac{r_s(z)}{z} - \frac{dr_s}{dz}. 
\end{equation}
Here $z$ measures the distance from the engine along the jet, while $r_s$ represents the shock radius (see {the left panel of} Fig.~\ref{Fig: Recollimation sketch}). 
Substituting the latter result into Eq.~\eqref{Eq: Pressure-balance recollimation} allows one to obtain an analytic solution of the first order differential equation for $r_s$, which reads
\begin{equation}
\label{Eq: shock radius}
    r_s(z) = \theta_0 (1+Az_*) z - A \theta_0 z^2,
\end{equation}
where $A = \sqrt{P_c(t) \pi u_0/[2 L_j]}$ and $z_*$ is the location, typically closer to the central engine, where the recollimation shock forms.
At this point it is possible to obtain the general expression for the angle $\psi$, $\sin{\psi}(z)= A \theta_0 z$, implying that $\psi$ ranges approximately between $\theta_0/2$ and $\theta_0$. 
The recollimation shock has a parabolic shape as shown in Fig.~\ref{Fig: Recollimation sketch}.
By setting $r_s = 0$ and $dr_s/dz = 0$ one can find, respectively, the position $\hat{z}= A^{-1} + z_* \approx A^{-1}$, where the recollimation shock converges on the jet axis, and the position $\bar{z} = \hat{z}/2$, where the shock surface is parallel to the jet axis and the jet changes from a conical to a cylindrical shape.  
The jet section beyond the recollimation shock can now easily be computed as 
\begin{equation}
\label{Eq: jet-surface}
    \Sigma_1(t) = \pi r_s^2(\bar{z}) = \pi \left( \frac{\hat{z}}{2} \right)^2 \theta_0^2 \approx \frac{L_j \theta_0^2}{2 P_c(t) u_0} \,.
\end{equation}

{\paragraph{Recollimation shock with Mach disk:}
If the cocoon pressure becomes sufficiently large, the small-angle approximation ($\psi \ll 1$) may break down, preventing a smooth recollimation of the flow. In this regime, the oblique shock may no longer be able to gradually deflect the jet material toward the axis, leading instead to the formation of a Mach disk near the recollimation point \citep[see, e.g.,][]{Perrucho2013,Mizuno_2015,Marti_2018}.
This scenario is particularly interesting because the resulting shock surface becomes nearly orthogonal to the jet bulk motion. As a consequence, the conversion of ram pressure into thermal pressure can become extremely efficient, potentially reaching ($k_p \sim \mathcal{O}(1)$), and may therefore provide better conditions for particle acceleration, especially in terms of fraction of ram pressure converted in cosmic-ray energy density. In this case, the recollimation shock structure is truncated at the Mach disk, as illustrated in the right panel of Fig.~\ref{Fig: Recollimation sketch}.
}

\subsubsection{Relevant scalings}

As the main focus of this work is on subrelativistic and mildly relativistic jets we proceed in this section under the physically motivated assumption of $\tilde{L} \ll 1$. 

Substituting Eqs.~\eqref{eq: cocoon pressure}-\eqref{Eq: jet-surface} into Eq.~\eqref{Eq: L-tilde definition}, it is possible to obtain the temporal evolution of $\tilde{L}$:
\begin{equation}
\label{eq: L-tilde vs time}
    \tilde{L}(t) = 
        \frac{4}{u_1^2} L_j^{2/5} \theta_0^{-8/5} \rho_{\rm A}^{-2/5} t^{-4/5}  \left[ \frac{(\gamma-1) \eta \alpha \beta^2 (u_0/u_1)^2}{2 \pi \zeta^2 (1+k_p)^{1/2} } \right]^{2/5}.
\end{equation}
Plugging this result in Eq.~\eqref{eq: cocoon pressure} the cocoon pressure reads
\begin{equation}
\label{eq: cocoon-pressure vs time}
    {P}_c(t) = 
        L_j^{2/5} \theta_0^{2/5} \rho_{\rm A}^{3/5} t^{-4/5}  \left[ \frac{(\gamma-1) \eta \alpha \beta^2 }{2 \pi (1+k_p)^{1/2}} \right]^{2/5} \left(\frac{u_1 \zeta}{u_0}\right)^{1/5}.
\end{equation}
According to Eq.~\eqref{eq: jet-head speed}, $v_h(t) \propto \tilde{L}^{1/2}(t) \sim t^{-2/5}$. Similarly, $v_c(t) \propto P_c^{1/2}(t) \sim t^{-2/5}$. 
In particular, one can see that $v_h/v_c \sim \theta_0^{-1}$.
This means that both the jet head and the cocoon side expand with the same self-similar index $\alpha = \beta = 3/5$ but with different normalization, in agreement with what is described in Appendix B.1.A of \citet{Bromberg_2011}.

From the latter results we can finally derive the expression for the jet head position as well as for the sideways cocoon extension and recollimation shock height:
\begin{align}
    &z_h(t) \approx 2 \, L_j^{1/5} \rho_{\rm A}^{-1/5} t^{3/5} \theta_0^{-4/5} \left[ \frac{\eta  (\gamma-1) (u_0/u_1)^2 (1+k_p)^2}{\pi \zeta^2} \right]^{1/5} \\
    &R_c(t) \approx L_j^{1/5} \rho_{\rm A}^{-1/5} t^{3/5} \theta_0^{1/5} \left[ \frac{\eta (\gamma-1) }{\pi (1+k_p)^{1/2}} \right]^{1/5} \left(\frac{u_1 \zeta}{u_0}\right)^{1/10} \\
    &\hat{z}(t) \approx 2 \, L_j^{3/10} \rho_{\rm A}^{-3/10} t^{2/5} \theta_0^{-1/5} u_0^{-1/2} \left[ \frac{(1+k_p) (u_0/u_1)}{\zeta \eta^2 \pi^3 (\gamma-1)^2} \right]^{1/10}.
    \label{eq: shock-lo}
\end{align}
We note that the jet head and the recollimation shock height share the same self-similar scalings of the astrophysical wind bubble \citep{Weaver_1977,Morlino2021}. We further discuss this aspect in Appendix~\ref{Subs: Appendix Bubble similarity}.

Finally, following \cite{Bromberg_2011}, we assume $\eta \approx 1$, 
which corresponds to a fully adiabatic cocoon valid when radiative losses are negligible.
If, on the contrary, energy losses are relevant, the effect of $\eta<1$ is to move the position of the recollimation shock farther away from the central object and, at the same time, to reduce the distance of the jet head, so that the size of the downstream is reduced. For example, assuming $\eta=0.01$ the recollimation shock moves farther by a factor of $\sim 2.5$ while the jet head gets closer by the same factor. We discuss the possible impact of this on the maximum energy of accelerated particles in Sect. \ref{subs: Parametric scan}.
{We stress that the dependence of the jet's parameters on the main physical quantities are fully consistent with those presented in \citet{Bromberg_2011}. 
The results presented in this section shall be considered as a subrelativistic extension of their theory where, differently from \citet{Bromberg_2011}, we include the possible dynamical impact of the recollimation shock strength through the parameter $k_p$.}

\section{Particle acceleration model}
\label{Sec: Transport}

Jets are extremely complex systems characterized by several plausible processes and locations able to accelerate particles. 
A nonexhaustive list includes DSA at shocks (e.g., forward, reverse, re-collimation, and internal), shear acceleration in the jet or at its borders \citep{Rieger2019}, magnetic reconnection \citep{Sironi2015} and turbulent acceleration (i.e., second-order Fermi process; see \citealt{Comisso2018,Matthews2020,Lemoine2018}, and references therein).
In nonrelativistic collimated jets, at large distances from the central engine where the flow is likely matter-dominated, DSA represents the primary particle acceleration mechanism. 
In this respect, jets are somewhat similar to wind bubbles.
In this work, we consider an idealized system focusing our attention on a jet at an advanced evolutionary stage that allows us to approximate the acceleration and transport of high-energy particles as steady in time.
We focus on the recollimation shock, hereafter labeled with the subscript \textit{sh}, as the most relevant site for DSA to take place, and we identify its position with $\hat{z}$ as described in Eq.~\eqref{eq: shock-lo}. 
We leave, instead, to a follow-up investigation the description of particle acceleration at the reverse and forward shocks at the jet head, given the complexity of particle dynamics in this region, determined by lateral advection and particle escape.

In what follows we develop an analytic modeling of the jet structure in Sect.~\ref{Subs: structure} and we provide a detailed description of our assumptions for the magnetic field and their impact on the diffusion coefficients. 
In Sect.~\ref{Subs: Transport} we describe our acceleration and transport model. In Appendix~\ref{Subs: Appendix Turbulence cascade} we provide analytic results describing different scenarios of turbulence decay in the downstream region.

\subsection{General properties of the system}
\label{Subs: structure}

The jet is launched by a central engine with terminal velocity $V_w$ and mass loss rate $\Dot{M}$.
As it travels along the launch direction, $z$, the outflow is also subject to lateral expansion. 
According to our dynamical model --- described in Sect.~\ref{Sec: dyn-and-structure} --- the jet expands conically with an opening angle $\theta_0$ up to the recollimation shock where it is focused along the direction of the axis. 
We parameterize the jet geometry in terms of a height-dependent area:
\begin{equation}
\label{Eq. jet section}
    A(z) = 
    \begin{cases}
        \pi z^2 \tan^2{\theta_0} \quad \quad z \lesssim z_{\rm sh} \\
        \pi z_{\rm sh}^2 \tan^2{\theta_0} \quad \quad z \gtrsim z_{\rm sh}
    \end{cases}.
\end{equation}
The behavior of the magnetic field in nonrelativistic and mildly relativistic jets, especially when far from the central engine, is highly uncertain. 
We thus assume subequipartition between the fluid ram pressure $\rho u^2$ and the magnetic field pressure $B^2/8 \pi$, parameterizing with $\epsilon_B$ the fraction of ram pressure converted into magnetic energy density. 
Finally, we assume the latter to be mostly of turbulent nature.
We stress that $\epsilon_B \ll 1$ so that the magnetic turbulence does not play a dynamically relevant role. 
We consider that the magnetic field has a typical coherence length $l_c$ and cascades toward smaller scales with a spectral index $\delta$. At the recollimation shock we consider a magnetic field compression by a factor of $\sqrt{11}$.

The parameterization of the magnetic field and the turbulence cascade allows us to characterize the diffusion coefficient as 
\begin{equation}
    D(z,p) = \frac{v(p)}{3}
    \begin{cases}
        r_L(z,p)^{2-\delta} l_c^{\delta-1} \quad r_L \leq l_c \\
        l_c [r_{L}(z,p) / l_c]^2 \quad r_L \geq l_c
    \end{cases},
\end{equation}
where we account both for the quasi-linear theory, valid for Larmor radii $r_L < l_c$, as well as the small pitch angle scattering regime taking place for $r_L > l_c$ \citep[see, e.g.,][]{Subedi2017,Dundovic2020}.  
In this work, we assume the jet diameter at the shock as the natural injection scale of the turbulence, i.e., $l_c \simeq 2\hat{z} \, {\rm tan} \theta_0$.
We focus on three typical scalings of the turbulence: Bohm ($\delta=1$), Kraichnan ($\delta = 3/2$) and Kolmogorov ($\delta=5/3$). 
The adoption of three different phenomenological diffusion models and their scalings provides insight into the possible outcomes of different diffusion theories, such as mirror diffusion \citep{Lazarian2021,BarretoMota2025}.
Finally, we account for the fact that turbulence may decay along the $z$-axis in the downstream, which, in turn, will reflect in an increasing diffusion coefficient. To this end, we explore different decay laws producing either power-law scaling, $D\propto z$ or $\propto z^2$, or exponential scaling, $D\propto e^{-z/N l_c}$, where $N$ is the number of coherence lengths after which the turbulence level becomes negligible.

\subsection{Acceleration and transport}
\label{Subs: Transport}

We consider only the transport along the jet axis direction in a system where the jet section may expand with altitude while we neglect lateral escape. 
This can be motivated by a suppressed diffusion perpendicular to the jet axis resulting from the typical topology of magnetic-field lines in jets, which are either aligned with the jet axis or wrapped around it. 
Nevertheless, the impact on the transport of a possible lateral escape as well as an energy loss term is discussed in Appendix~\ref{Subs: Appendix sink term}.
We thus model the system adopting the one-dimensional stationary transport equation:
\begin{align}
    \label{Eq. Transport}
    A(z) & u(z) \frac{\partial f(z,p)}{\partial z} = \frac{\partial }{\partial z} \left[ A(z) D(z,p) \frac{\partial f(z,p)}{\partial z} \right] + \nonumber \\
    + \frac{1}{3} & p \frac{\partial f(z,p)}{\partial p} \frac{\partial [A(z)u(z)]}{\partial z} + A(z) Q(z,p),
\end{align}
where $f$ is the phase space density of accelerated particles which depends on the spatial  coordinate $z$ and momentum $p$. $u$ is the plasma speed, $D$ the diffusion coefficient and $A$ is the z-dependent function describing the expansion of the wind transverse section. 
The injection term $Q$ reads as 
\begin{equation}
    \label{Eq. Injection}
    Q(z,p) = Q_{\rm sh}(p) \delta[z-z_{\rm sh}] = \eta_{\rm eff} \frac{n_0 u_{0,n}}{4 \pi p^2} \delta[p-p_{\rm inj}] \delta[z-z_{\rm sh}] ,
\end{equation}
where $n_0$ and $u_{0,n}$ are the gas density and wind speed in the immediate upstream of the wind termination shock, while $\eta_{\rm eff}$ is the fraction of all upstream thermal protons getting injected in the DSA process. 
Note that the upstream wind velocity at the shock  $u_0$ coincides with the terminal wind speed, while the projection along the shock normal is $u_{0,n} = u_0 \sin \psi$.
The injection is localized at the termination shock position, $z_{\rm sh}$, and the injection momentum is $p_{\rm inj}=m_p c$. 
We also note that lower injection momenta would not have a relevant impact as their contribution to the pressure at the shock would be subdominant.

In agreement with the cylindrical approximation, the velocity profile is assumed as
\begin{equation}
    u(z) = u_0(z) + [u_1(z) - u_0(z)] \, \theta[z-z_{\rm sh}],
\end{equation}
where the upstream velocity, $u_0$, is connected to the downstream one, $u_1$, through the jump conditions typical of strong shocks, namely $u_{0,n} = r u_{1,n} $, where $r$ is the compression ratio. Here we assume the value $r=4$ of strong shocks.
We stress that the jump condition holds for the component along the shock normal, whereas the tangential one is conserved, $u_{0,t} = u_{1,t}$.
For the upstream wind we assume a fast rising profile that approaches 0 at $z=0$. As the wind launching region is much smaller than the typical diffusion length of particles able to propagate upwind, the details of the wind profile for $z \longrightarrow 0$ do not have an impact on our results. 
Therefore, for simplicity we parameterize the upstream wind adopting a ${\rm tan^{-1}}(z)$ profile that quickly converges to the terminal wind speed $u_0$. 
In the downstream region, we assume a constant wind profile $u_1(z) = u_1$.
The gas number density at the shock is defined consistently with the jump conditions of the normal component of the wind velocity.
The gas profile, the jet area, and the wind speed obey the mass conservation equation 
\begin{equation}
    \label{Eq. mass cons}
    A(z) \, \rho(z) \, u(z) = \Dot{M},
\end{equation}
where $\rho = n m_p$ is the gas mass density in the jet.
Consequently, the knowledge of $A$ and $u$ sets also the gas density profile.

\subsubsection{Upstream}
\label{Subs: Upstream}

To find the general solution of Eq.~\eqref{Eq. Transport} we adopt an iterative technique similar to the one described in \cite{Morlino2021}, \cite{Peretti_2022,Peretti_2023,Peretti_2025_ULX}, and \cite{Payel_2023}. 
For $z<z_{\rm sh}$ we indicate $f(z_{\rm sh},p)=f_{\rm sh}(p)$, while we assume that the system is symmetric with respect to the plane $z=0$. 
This leads to a null net flux condition at $z=0$ which, for realistic wind profiles characterized by no discontinuities, namely $u(0)=0$, leads to the condition $\partial_z f|_{z=0}=0$. 

We approach the upstream solution by integrating Eq.~\eqref{Eq. Transport} from $z_{\rm min}=0$ up to a generic $z<z_{\rm sh}$. 
The result is 
\begin{equation}
    \label{Eq. Up1}
    A(z) D_0(z,p) \partial_z f_0(z,p) = A(z)u_0(z) f_0(z,p) + G_0(z,p),    
\end{equation}
where the subscript $0$ refers to upstream parameters and
\begin{align}    
    \label{Eq. G1}
   &G_{0}(z,p) = \frac{1}{3} \int_0^{z} dz' f_0(z',p) \partial_{z'}[A(z')u_0(z')] g_0(z',p), \\
    \label{Eq. zeta_1}
    &g_0(z,p) =  - \frac{{\rm dln} \, p^3f_0(z,p)}{{\rm dln} \,p} \,.
\end{align}
Physically, Eq.~\eqref{Eq. Up1} states that the balance between the diffusing flux, $-AD\partial_z f$, and the advected flux, $Auf$, is broken by the function $G_0$, which accounts for adiabatic losses/gain, the geometry of the system and its size. 
In Eq.~\eqref{Eq. Up1} one can recognize a total derivative with respect to the spatial coordinate, so that it can be rewritten as 
\begin{equation}
\label{Eq. Up2}
    \partial_z \left\{ f_0(z,p) \exp\left[ \int_{z}^{z_{sh}} dz' \frac{u_{\rm eff,0}(z',p)}{D_0(z',p)} \right] \right\}  = 0 ,
\end{equation}
where the effective velocity $u_{\rm eff,0}$ is defined as follows:
\begin{equation}
    \label{Eq. Eff. up vel.}
    u_{\rm eff,0}(z,p) = u_0(z) \left[ 1 + \frac{G_0(z,p)}{A(z)u_0(z)f_0(z,p)} \right]. 
\end{equation}
Requiring that the solution is continuous at the recollimation shock, namely $f_0(z_{\rm sh},p) = f_{\rm sh}(p)$, it is straightforward to obtain the general solution to Eq.~\eqref{Eq. Up2}:
\begin{equation}
    \label{Eq. Up-Sol}
    f_0(z,p) = f_{\rm sh}(p) \exp{\left[ -\int_z^{z_{\rm sh}} dz' \frac{u_{\rm eff,0}(z',p)}{D_0(z',p)} \right]}.
\end{equation}
In this upstream solution one can recognize the same behavior of the infinite planar shock solution that gets exponentially suppressed with distance from the shock.

\subsubsection{Downstream}
\label{Subs: Downstream}

For the downstream region, as boundary conditions we require continuity of the particle distribution function at the recollimation shock, namely $f_1(z_{\rm sh},p)=f_{\rm sh}(p)$, while we assume a free escape boundary at the jet head $f(z_{\rm esc},p) = 0$. 
{While we comment in Sect.~\ref{Sec: discussion} on the possibility of re-acceleration at the jet head, the adoption of such a free escape boundary condition at that location reflects the fact that particles are unlikely to return to the recollimation shock, and thus cannot further participate in the acceleration process.}

In analogy with the upstream region, we perform a spatial integration of Eq.~\eqref{Eq. Transport} from $z>z_{\rm sh}$ up to $z_{\rm esc}$. The result reads
\begin{align}
    \label{Eq. Down1}
    A(z)& D_1(z,p) \partial_z f_1(z,p) - A(z)u_1(z) f_1(z,p) \, + \nonumber \\ 
    & + G_1(z,p) = - A(z_{\rm esc}) j_{\rm esc}(p),
\end{align}
where the subscript 1 refers to the quantity evaluated in the downstream region and $j_{\rm esc} = - D_1 \partial_z f|_{z=z_{\rm fs}}$ is the escaping flux.
The function $G_1$ appearing in Eq.~\eqref{Eq. Down1} reads
\begin{align}
    &G_1(z,p) = \frac{1}{3} \int_z^{z_{\rm esc}} dz' \, f_1(z',p) \partial_{z'}[A(z')u_1(z')] g_1(z',p) \\
    &g_1(z,p) = - \frac{{\rm dln} \, p^3f_1(z,p)}{{\rm dln} \,p}.
\end{align}
We note that, qualitatively, Eq.~\eqref{Eq. Down1} has the identical physical interpretation of Eq.~\eqref{Eq. Up1}. The main difference here is that, in addition to the adiabatic term, an escape term also appears.
Similarly to the upstream region, it is possible to recognize in Eq.~\eqref{Eq. Down1} a total derivative. Performing another spatial integration and adopting the continuity condition at the recollimation shock, one obtains the following:
\begin{align}
    \label{Eq. Down2}
    f_1 &(z,p) \exp\left[-\int_{z_{\rm sh}}^{z}dz' u_{\rm eff,1}(z',p)/D_1(z',p)\right] = \nonumber \\ 
    & = f_{sh}(p) - A(z_{\rm esc}) j_{\rm esc}(p) \mathcal{I}(z,p),
\end{align}
where 
\begin{align}
    &\mathcal{I}(z,p) = \int_{z_{\rm sh}}^{z} dz' \frac{\exp\left[- \int_{z_{\rm sh}}^{z'} dz'' u_{\rm eff,1}(z'',p)/D_1(z'',p) \right] }{A(z')D_1(z',p)}, \\
    &u_{\rm eff,1}(z,p) = u_1(z) \left[ 1 - \frac{G_1(z,p)}{A(z)U_1(z)f_1(z,p)} \right].
\end{align}

Applying the free escape boundary condition, $f(z_{\rm esc}, p) =0$, to Eq.~\eqref{Eq. Down2} we get the expression of the escaping flux:
\begin{equation}
    \label{Eq. Esc-flux}
    j_{\rm esc}(p) = f_{\rm sh}(p) \frac{1}{A(z_{\rm esc}) \mathcal{I}(z_{\rm esc},p) } .
\end{equation}
Finally, the downstream solution can be obtained combining Eqs.~\eqref{Eq. Down2} and \eqref{Eq. Esc-flux}, which gives
\begin{equation}
    \label{Eq. Down-Sol}
    f_1(z,p) = f_{\rm sh}(p)  \exp\left[\int_{z_{\rm sh}}^z dz' \frac{u_{\rm eff,1}(z',p)}{D_1(z',p)} \right]  \left[ 1 - \frac{\mathcal{I}(z,p)}{\mathcal{I}(z_{\rm esc},p)}  \right]  .
\end{equation}

Interestingly, as detailed in Appendix~\ref{Subs: Appendix Turbulence cascade}, the downstream solution acquires a simple analytic form for the cases of turbulence decay of our interest, as described in Sect.~\ref{Subs: structure}.

\subsubsection{Recollimation shock}
\label{Subs: Shock}

We assume that the fluid experiences an average angle of impact with the shock surface $\psi_{\rm eff} = 3 \theta_0/4$, obtained mediating Eq.~\eqref{Eq: proxi angle} between $\bar{z}$ and $\hat{z}$. 
These approximations allow us to preserve the 1D description of the acceleration and transport model. 

The solution at the termination shock is found integrating the transport equation in an infinitely small spatial range around the termination shock, namely from $z_{\rm sh}^-$ up to $z_{\rm sh}^+$. 
The result of the integration, leads to a projection along the shock normal which reads
\begin{align}
    0 = [\sin \psi^* D \partial_z f]|^{z_{\rm sh}^{+}}_{z_{\rm sh}^{-}} - \frac{(u_{0,n}-u_{1,n})}{3} p \partial_p f_{\rm sh}(p) + Q_{\rm sh}(p)  , 
    \label{Eq. shock1}
\end{align}
where $\psi^*$ is the angle between the shock surface and the fluid velocity vector. 
In particular, $\psi^*=\psi_{\rm eff}$ in the upstream region, while its value in the downstream is found consistently with the compression of the velocity along the shock normal.
In the adiabatic term, we used the identity $\partial_z \theta[z-z_{\rm sh}] = \delta[z-z_{\rm sh}] $ and the advective term (left hand side) goes to 0 after an integration by parts.
Substituting in Eq.~\eqref{Eq. shock1} the expressions of Eqs.~\eqref{Eq. Up1} and \eqref{Eq. Down1} one obtains the following: 
\begin{align}
    \frac{s Q_{\rm sh}(p)}{u_{0,n} p} & = \partial_p f_{\rm sh}(p) + \frac{s}{p} f_{\rm sh}(p) + \frac{s}{p} \frac{G_0(z_{\rm sh},p)}{A(z_{\rm sh})u_0 f_{\rm sh}(p)} f_{\rm sh}(p) + \nonumber \\
    + & \frac{s}{r p} f_{\rm sh}(p) \left[ \frac{G_1(z_{\rm sh},p)}{u_1 A(z_{\rm sh}) f_{\rm sh}(p)} + \frac{1}{u_1 A(z_{\rm sh}) \mathcal{I}(z_{\rm esc}) } - 1 \right],
\end{align}
where $s=3u_{0,n}/(u_{0,n}-u_{1,n})$.
In the latter expression it is possible to recognize a total derivative so that it can be rewritten as
\begin{equation}
\label{Eq. shock2}
    \partial_p[f_{\rm sh}(p)p^s e^{\tilde{\Gamma}_0(p)} e^{\tilde{\Gamma}_1}(p)] = \frac{s}{u_{0,n}} \frac{Q_{\rm sh}(p)}{p} p^s e^{\tilde{\Gamma}_0(p)} e^{\tilde{\Gamma}_1(p)},
\end{equation}
where the functions appearing in the exponential are defined as 
\begin{align}
    &\tilde{\Gamma}_0(p) = s \int_{p_0}^p \frac{dp'}{p'} \frac{G_0(z_{\rm sh},p')}{A(z_{\rm sh})u_0 f_{\rm sh}(p')}  \\
    &\tilde{\Gamma}_1(p) = \frac{s}{r} \int_{p_0}^p \frac{dp'}{p'} \left[ \frac{G_1(z_{\rm sh},p)}{u_1 A(z_{\rm sh}) f_{\rm sh}(p)} + \frac{1}{u_1 A(z_{\rm sh}) \mathcal{I}(z_{\rm esc}) } - 1 \right] ,
\end{align}
where $p_0$ is an arbitrarily low value in the momentum scale smaller than $p_{\rm inj}$.
It is now possible to integrate \eqref{Eq. shock2} and obtain the following result for the solution at the termination shock,%
\begin{equation}
    \label{Eq. Shock solution}
    f_{\rm sh}(p) = \eta_{\rm eff} \frac{s n_0}{4 \pi p_{\rm inj}^3} \left( \frac{p}{p_{\rm inj}} \right)^{-s} e^{-\Gamma_0(p)} e^{-\Gamma_1(p)}, 
\end{equation}
where $\Gamma_{0,1}(p) = \tilde{\Gamma}_{0,1}(p) - \tilde{\Gamma}_{0,1}(p_{\rm inj})$.

The general solution of the transport equation is found with an iterative algorithm that can be found in Appendix~\ref{Subs: Appendix algorithm}. Finally, in Appendix~\ref{Appendix: analytic considerations} we discuss some analytic aspects of the general solution and possible similarities with the wind termination shock in wind bubbles.

\section{Results}
\label{Sec: Results}

\begin{table*}[t]
\caption{Prototypical sources and maximum energies.}
\label{Tab: prototypes}
\centering
\begin{tabular}{l|llllll|lll}
\hline
 &  $\dot{M} \, \rm [M_{\odot} \,yr^{-1}]$ & $u_0 \, \rm [km \,s^{-1}]$ & $\theta_0$  & $n_A \, \rm [cm^{-3}]$ & $T_{\rm age} \, \rm [yr]$ & $\epsilon_B$ & $E_{\rm max} \, [\delta=5/3]$ & $E_{\rm max} \, [\delta=3/2]$ & $E_{\rm max} \, [\delta=1]$   \\
\hline
SEY$_0$ &  $10^{-2}$ & $3\cdot 10^4$ & $10^{\circ}$  & $10^3$ & $10^5$ &  $10^{-5}$ & 6 {\rm PeV} & 50 PeV & 240 PeV \\
MQ$_0$ &  $10^{-6}$ & $ 3\cdot 10^4$ & $2^{\circ}$ & $1$ & $10^4$ &  $10^{-5}$  & 5 {\rm PeV} & 7 PeV & 8 PeV  \\
PS$_0$ & $10^{-7}$ & $5 \cdot 10^2$ & $5^{\circ}$ & $10^4$ & $10^4$ &  $10^{-5}$ & 2 {\rm GeV} & 28 GeV & 4 TeV  \\
\hline
\end{tabular}
\tablefoot{Prototypical protostellar jets parameters ($PS_0$) are consistent with HH80-81 as discussed in \cite{Rodriguez-Kamentzky2017}, while those of the prototypical microquasar ($MQ_0$) are consistent with SS433 as presented in \cite{Fabrika_2004}. The parameters inferred for the radio jet in NGC~4151 \citep{Ulvestad_2005,Willams_2017} are assumed as prototypical for jets in Seyfert galaxies ($SEY_0$).}
\end{table*}

Even though they have qualitative similarities in the structure and the dynamics, the power content and the size of jets from protostars, microquasars and active galaxies belong to completely different classes. 
In light of this, we found it practical to set one reference setup as a prototype source for each of the above mentioned classes.
In Table~\ref{Tab: prototypes} we report the relevant parameters assumed with the computed maximum energies for the accelerated particles. 
For each class, we explored different diffusion coefficients, from the most efficient in scattering, Bohm ($\delta=1$), to the least efficient, Kolmogorov ($\delta=5/3$). 
As its properties lie somewhat in between, we adopt the Kraichnan diffusion coefficient ($\delta=3/2$) as a benchmark. In addition, for simplicity, we adopt as a reference, the scenario of spatially constant downstream diffusion coefficient (see Appendix~\ref{Subs: Appendix Turbulence cascade}), while we discuss in Sect.~\ref{subs: Parametric scan} the impact of magnetic turbulence decay in the downstream.
Table~\ref{Tab: prototypes} also reports the main references where such parameter values can be retrieved. 
In the following, we discuss the different source classes in Sects.~\ref{subs: Seyfert}-\ref{subs: prototype PS}. 
We delve into the maximum energy with analytic estimates in Sect.~\ref{Sec: E_max}.
In Sect.~\ref{subs: Parametric scan} we evaluate the dependence of particle spectra on the main parameters. We finally comment on the multi-messenger emission from these systems in Sect.~\ref{subs: comments}.

\subsection{Prototype Seyfert jet}
\label{subs: Seyfert}
\begin{figure}[t]
    \centering
    \includegraphics[width=1\columnwidth]{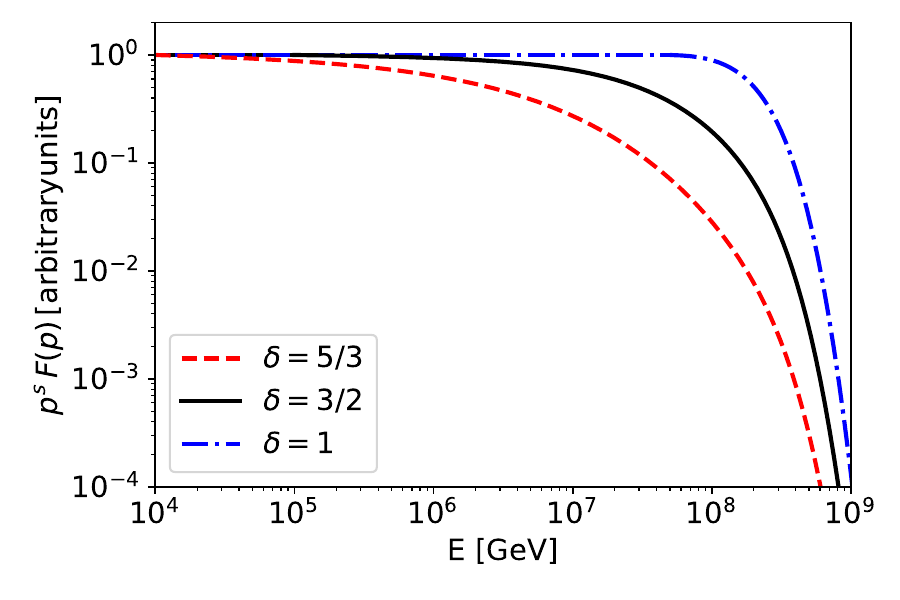}
    \caption{Seyfert prototype.
    Spectra of accelerated particles at the recollimation shock. The dot-dashed blue curve represents the Bohm scenario, while the solid black and the dashed red represent the Kraichnan and Kolmogorov cases, respectively.}
    \label{Fig: SEY}
\end{figure}
{Although radio-loud narrow-line Seyfert 1 galaxies represent an important exception~\citep{Foschini_2020},}
Seyfert jets are {generally} among the weakest and slowest collimated outflows launched by accreting supermassive black holes in active galaxies.
They typically feature a large variability in mass loss rate and mildly relativistic wind speed. 
The prototypical kinetic power of these jets, around $L_{\rm kin}=10^{42}\, \rm erg \, s^{-1}$, is inferred for both NGC~1068 \citep[see][and references therein]{Padovani_24_1068} and NGC~4151 \citep{Ulvestad_2005,Willams_2017}. 
To our knowledge, the opening angle of weak radio jets in Seyfert is not well constrained,  so we assume $\theta_0 = 10^{\circ}$ as a benchmark. 
In the central parsecs of Seyfert galaxies the interstellar medium is significantly denser than the average one. Therefore, we adopt an effective density of $n_A = 10^3 \, \rm cm^{-3}$ \citep{Sanchez-Garcia2022}. 
We set the typical age of Seyfert jets to $T_{\rm age}=10^5 \, \rm yr$, which corresponds to a fraction $1-10\%$ of the typical AGN duty cycle. 
Finally, we assume a typical value of $\epsilon_B=10^{-5}$, which allows the recollimation shock to remain super-Alfvénic while maintaining magnetic field strengths in the typical range of 0.1–1 mG, as inferred from radio knots observations \citep{Gallimore96_1,Willams_2017}.
The reference set of parameters for the prototypical Seyfert jet (SEY$_0$) are summarized in Table~\ref{Tab: prototypes}.

\begin{figure}[t]
    \centering
    \includegraphics[width=1\columnwidth]{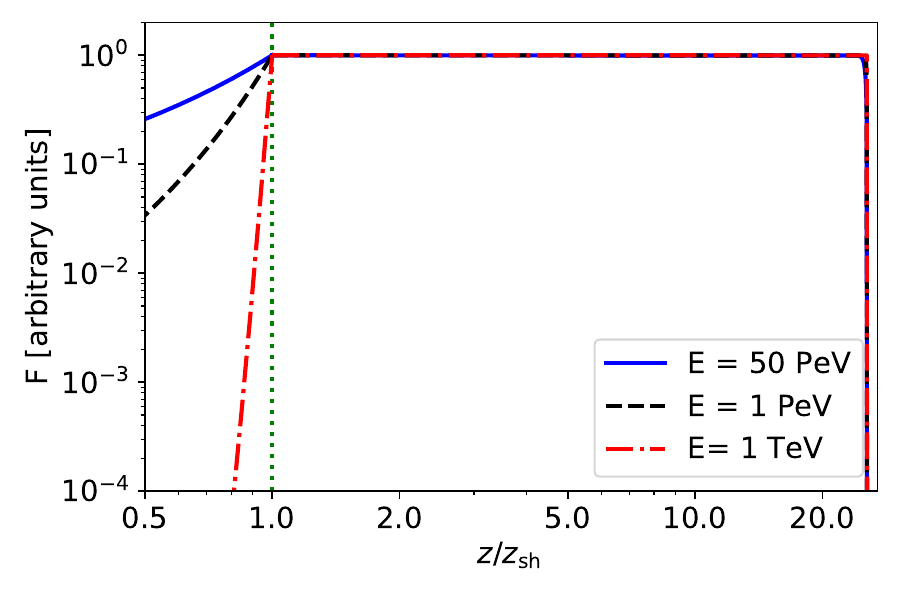} \quad \includegraphics[width=1\columnwidth]{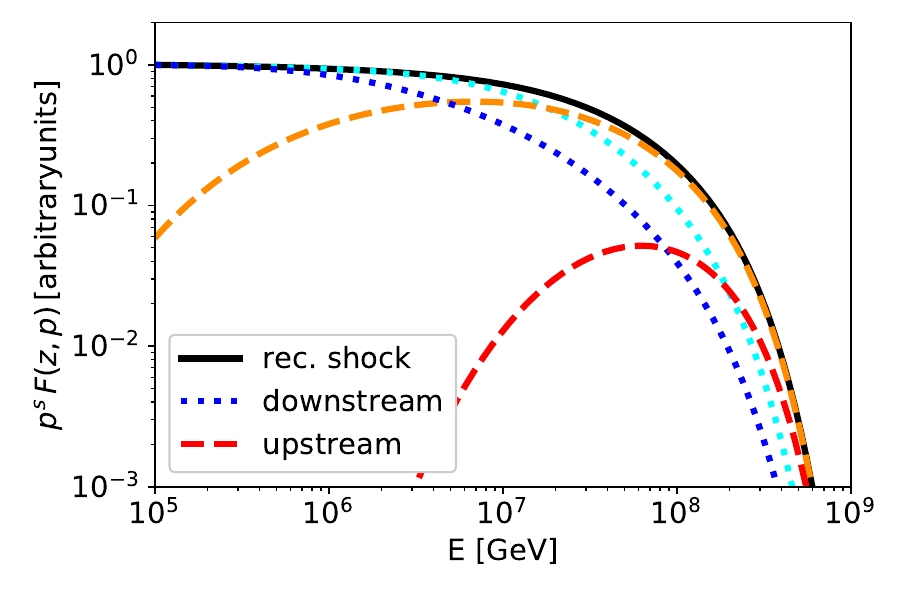}
    \caption{Radial behavior of the solution for the case of Seyferts jets. \textit{Top}: Particle distribution along $z$ computed at different energies. The vertical dotted line represents the recollimation shock location, while the jet head is located at approximately 20 times $z_{\rm sh}$. \textit{Bottom}: Solution computed at different positions in the jet. The solid black line represents the solution at the shock position;  the dashed lines refer to the upstream (orange for $z= 0.97 z_{\rm sh}$ and red for $z = 0.3 z_{\rm sh}$) and the dotted lines to the downstream (cyan and blue for $z \gtrsim 0.98 \, z_{\rm esc}$ and $z \gtrsim 0.99 \, z_{\rm esc}$, respectively).}
    \label{Fig: Radial explanation}
\end{figure}

Figure~\ref{Fig: SEY} illustrates the spectra of the solutions at the recollimation shock. 
Bohm diffusion   reaches the highest energy, $E_{\rm max} \sim 240 \, \rm PeV$, while Kraichnan and Kolmogorov diffusion   reach approximately $\sim 50$ PeV and a few PeV, respectively.
One can generally conclude that, depending on the turbulence scenario, Seyfert jets can accelerate particles from PeV up to the EeV range, where the latter can be reached under optimistic parametric assumptions (e.g., small opening angle, high kinetic power and/or Bohm diffusion).
It is interesting to note that, while at low energy all turbulence types predict a power law $p^{-s}$, the cutoff at higher energies has different shapes, ranging from sharp for the Bohm case to shallower for the Kolmogorov case, with the Kraichnan case falling in between. 
Such a behavior is analogous to what is expected from DSA models in stellar wind bubbles
\citep{Morlino2021,Peretti_2022,Peretti_2023,Peretti_2025_ULX,Payel_2023}.

\begin{figure}[t]
    \centering
   \includegraphics[width=1\columnwidth]{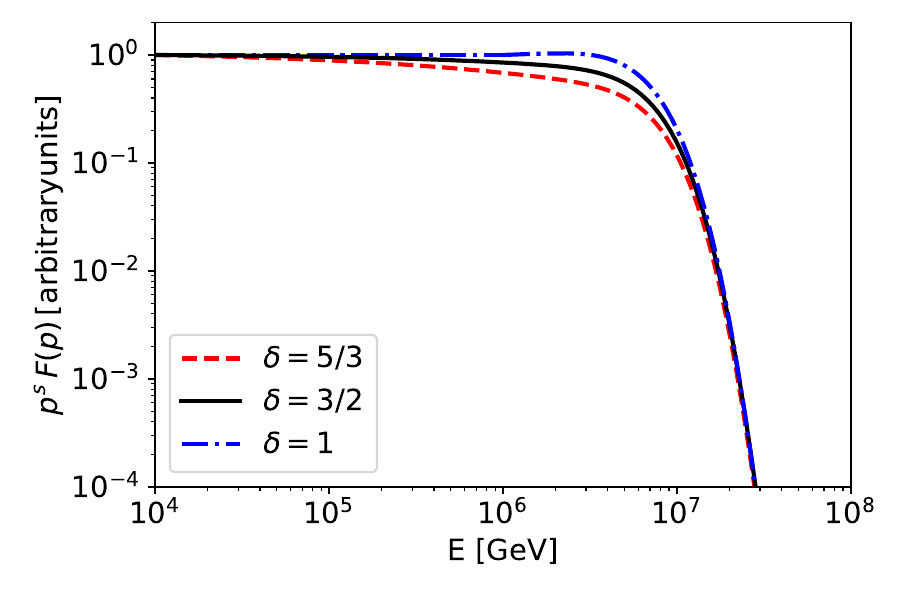}
    \caption{Microquasar prototype. 
    Particle spectra at the recollimation shock. The linestyles and colors are identical to Fig.~\ref{Fig: SEY}.}
    \label{Fig: MQ}
\end{figure}

Figure~\ref{Fig: Radial explanation} illustrates how particles behave in the system for the case of prototype Seyfert ($\delta=3/2$). The top panel illustrates the particle distribution as a function of altitude --- in the jet --- at different energies. The upstream region exhibits an exponential decay away from the shock, regulated by the diffusion length $D_0/u_0$, as is typical in DSA. In particular, only the highest energy particles can propagate sufficiently far upstream against the jet flow. The downstream region is almost completely dominated by advection. This results in the approximately homogeneous distribution, from the recollimation shock up to the jet head.
The bottom panel illustrates the behavior of particle spectra throughout the jet. Dashed lines correspond to locations in the upstream, while dotted lines to locations in the downstream. 
The upstream spectra clearly show that only the highest energy particles propagate against the jet flow, while the downstream spectra feature the softening due to diffusion while approaching the escape surface at the jet head.

\subsection{Prototype microquasar jet}
\label{subs: prototype MQ}

Inspired by the mildly relativistic velocities inferred for SS433 \citep{Marshall2002},
we adopt a conservative jet speed of the order of 0.1 c and a mass loss rate comparable with the estimated total kinetic power \citep{Fabrika_2004}. 
We assume a typical opening angle of $\theta_0=2^{\circ}$ and an age of $10^4$ yrs \citep[][]{Fabrika_2004}.
Finally, we consider an external medium density comparable with the one of the Galactic interstellar medium, $n_A \approx 1 \, \rm cm^{-3}$, and we assume a value for $\epsilon_B$ ($10^{-5}$) that allows for a magnetic field of the order of a few $10 \, \mu \rm G$, compatible with measurements of knots in SS433 \citep{Tsuji_2025}, and guaranteeing the shock to be super-Alfv\'enic. 
The parameters for the prototypical microquasar (MQ$_0$) can be found in Table~\ref{Tab: prototypes}. 

Differently from the Seyfert case, the maximum energy is approximately falling in the transition region to weak diffusion ($r_L \approx l_c$), where all types of diffusion behave approximately in the same way, resulting in a similar maximum energy. 
This can be clearly seen in Fig.~\ref{Fig: MQ}, where the spectra resulting from the three diffusion scenarios are compared. 
As one can also see from Table~\ref{Tab: prototypes}, the range of maximum energies varies within less than a factor of 2, from $\sim 5$ to $\sim 8$ PeV. 

\subsection{Prototype protostellar jet}
\label{subs: prototype PS}

\begin{figure}[t]
    \centering
   \includegraphics[width=1\columnwidth]{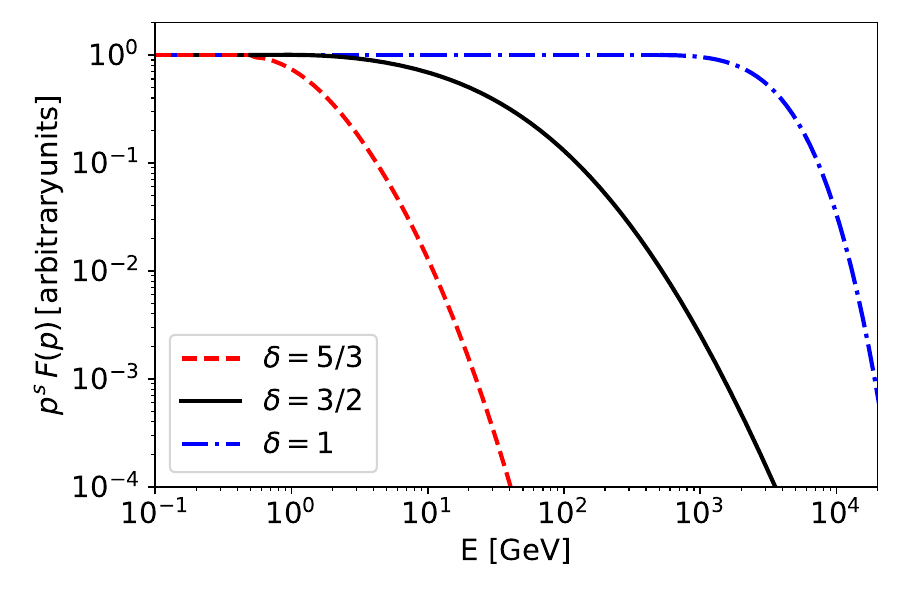}
    \caption{Protostar prototype. 
    Particle spectra at the recollimation shock. The linestyles and colors are identical to Fig.~\ref{Fig: SEY}.}
    \label{Fig: PS}
\end{figure}

Protostellar jets are the slowest class of astrophysical collimated jets. Their typical velocity ranges from about $10^2 \, \rm km \, s^{-1}$ up to $10^3 \, \rm km \, s^{-1}$ and their mass loss rate can exceed $10^{-7} \, \rm M_{\odot} \, yr^{-1}$ \citep[see, e.g.,][]{Rodriguez-Kamentzky2017,Araudo2021_protostar}. The jet half-opening angle is expected to range from a few degrees up to ten degrees, and  the period of activity can extend for several tens of kilo-years \citep{Frank_2014}. As density of the surrounding medium we adopt $n_A=10^4 \, \rm cm^{-3}$, which is typical of giant molecular clouds, where protostellar system are likely to be found. Finally, we assume $\epsilon_B=10^{-5}$, which guarantees that the shock is super-Alfv\'enic with a magnetic field strength of $B\approx 40 \, \mu G$. Such a value agrees with the magnetic field expected at a distance of $0.1$ pc from the protostar if the field at its surface is of the order of mG to G.
The parameters for the prototypical protostellar jet (PS$_0$) are summarized in Table~\ref{Tab: prototypes}.

As done for the previous source classes, in Fig.~\ref{Fig: PS} we show the spectra of accelerated particles at the recollimation shock.
It is possible to see that the maximum energy changes significantly between different diffusion scenarios. In particular, Kolmogorov diffusion seems unable to efficiently energize protons above a couple of GeV, while Kraichnan and Bohm diffusion   reach a few tens of GeV and a few TeV, respectively.
The computed values of the maximum energies for the different diffusion scenarios are reported in the last columns of Table~\ref{Tab: prototypes}.
We  also note that, given the typical size of the system of $\sim 0.1-1$ pc, energy losses due to Coulomb scattering and ionization as well as proton-proton (pp) interactions at high energy do not play a relevant role on the scales of our interest, namely at the location of the recollimation shock. 

\begin{figure}[t]
    \centering
    \includegraphics[width=1\columnwidth]{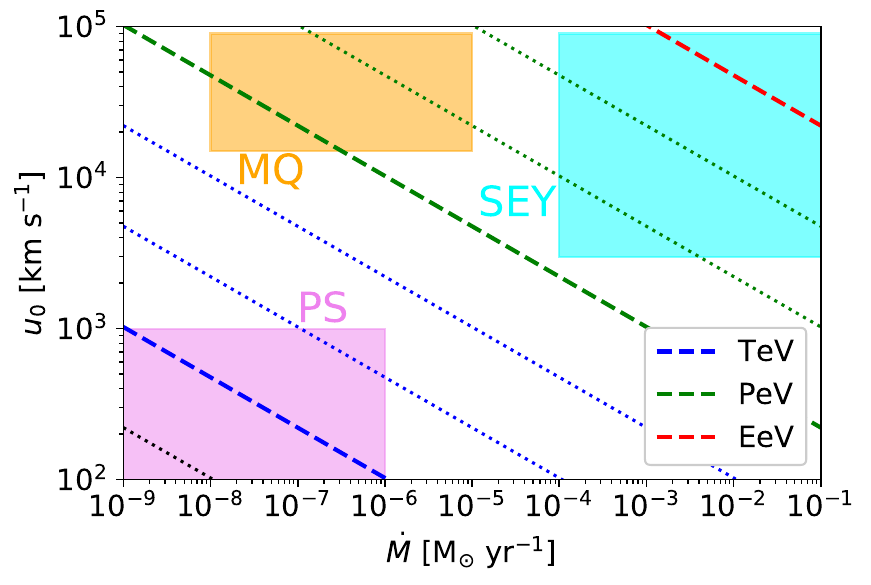}
    \caption{Mass-loss rate to terminal wind speed ($\dot{M}-u_0$) parametric plot of the maximum energy achievable by the subrelativistic jet classes analyzed in this work assuming the typical opening angle of $4^{\circ}$ for all three source classes. PS, MQ, and SEY respectively stand  for protostar (pink), microquasar (orange), and Seyfert (cyan). The diagonal lines indicate the different decades in maximum energy. The energy increases from bottom left ($10^2$ GeV) to top right (EeV). The dashed lines represent respectively TeV, PeV, and EeV, while the dotted lines are intermediate decades.}
    \label{Fig: Hillas jet}
\end{figure}

\subsection{Qualitative considerations on the maximum energy}
\label{Sec: E_max}

The maximum energy of particles accelerated at the recollimation shock of jets is one of the main results of this work.
Despite the fact that there are several parameters that influence the maximum energy, it is very useful to discuss qualitative estimates that are naturally set by the size of the system. 
In particular, the upstream region, compared to the downstream one, can be expected to set the most stringent conditions unless the turbulence decays extremely rapidly after the recollimation shock (see Sect. \ref{subs: Parametric scan} for a detailed discussion).

An estimate of the maximum energy can be obtained assuming that the upstream diffusion length of particles at the maximum energy, $D(E_{\rm max})/u_0$, is as large as the upstream region itself, $z_{\rm sh}$.
Here $u_0$ identifies the upstream velocity.
The latter condition is the analog of the Hillas criterion \citep{Hillas1984}, and in order to obtain an upper limit, we also assume Bohm diffusion. 
Recalling that the magnetic field pressure can be parameterized as a fraction $\epsilon_B$ of the ram pressure, we obtain the following general analytic expression:
\begin{equation}
\label{eq: Hillas}
    {E_{\rm max}} \lesssim 3 \left( \frac{\epsilon_B}{10^{-5}} \right)^{1/2}
    \left( \frac{\dot{M}}{10^{-6} \rm M_{\odot} \, yr^{-1}} \right)^{1/2}
    \left( \frac{u_0}{0.1 \, c} \right)^{3/2} \left( \frac{\theta_0}{4^{\circ}} \right)^{-1} {\rm PeV} 
.\end{equation}
It is interesting to note that the dependence of $E_{\rm max}$ on the mass-loss rate $\dot{M}$ and on the velocity $u_0$ is identical to the wind bubble scenario \citep{Peretti_2025_ULX}, whereas in the case of jets, there is an additional dependence on the inverse of the jet opening angle. The latter dependence is crucial for allowing jets to reach extremely high energies while the magnetic field can be dynamically subdominant.

We observe that, in general, the combination of Bohm diffusion and the equality between the diffusion length and the size of the upstream should set an upper limit to the maximum energy. However, we point out that this is not necessarily the case in the context of our model. In fact, high-energy particles that are propagating sufficiently far in the upstream region experience a higher magnetic field and, in turn, a smaller diffusion coefficient---this is a natural result of our parameterization of the magnetic field, $P_B = \epsilon_B P_{\rm ram}$. Such a space-dependence of the magnetic field can result in maximum energies slightly exceeding the estimate provided in Eq.~\eqref{eq: Hillas}, without resulting in a violation of the Hillas criterion.

Figure~\ref{Fig: Hillas jet} compares the maximum energy of the three jet classes analyzed in this work as a function of $\dot{M}$ and $u_0$ adopting the same opening angle $\theta_0=4^{\circ}$. 
It is possible to observe that protostellar jets are limited to the GeV-TeV energy domain; Microquasars can be expected to accelerate particles from hundreds of TeV up to tens of PeV for the most extreme sources, while Seyfert jets can accelerate particles from PeV up to the EeV domain. 

We finally point out that, while the scaling with $\dot{M}$ can be expected for all the considered diffusion scenarios, the scaling $E_{\rm max} \sim u_0^{3/2}$ is a result limited to Bohm diffusion. 
For a generic turbulence slope of index $\delta$ one can expect the dependence on the velocity to become  of the order $E_{\rm max} \sim u_0^{(4-\delta)/(4-2\delta)}$ with some possible fluctuations due to the dependence of $E_{\rm max}$ also on $z_{\rm sh}^{(\delta-1)/(2-\delta)}$, where $z_{\rm sh}$, in turn, mildly depends on $u_0$ with a power law index $0.1$. 
We additionally observe a possible dependence of the maximum energy on the coherence length of the magnetic field as $E_{\rm max}\sim l_c^{(1-\delta)/(2-\delta)}$.   
In practice, we adopt a physically motivated reference value for $l_c$ and, in the absence of additional relevant scales in the problem, we do not treat it as a free parameter. Nevertheless, in the following section we explore the possibility of having a somewhat smaller coherence length.

\subsection{Parameter space scan}
\label{subs: Parametric scan}

\begin{table}[t]
\caption{Parameter space scan.}
\begin{tabular}{l|lll}
\hline
 &  $E_{\max}$ (PS) & $E_{\max}$ (MQ) & $E_{\max}$ (SEY) \\
\hline \hline
prototype & 28 {\rm GeV} & 7 {\rm PeV} & 50 {\rm PeV} \\
\hline
$2 \, u_0$ & 150 {\rm GeV} & 18 {\rm PeV} & 278 {\rm PeV} \\
$ 1/2 \,  u_0$ & 8 {\rm GeV} & 2 {\rm PeV} & 9 {\rm PeV} \\
$ 10 \,  \dot{M}$ & 80 {\rm GeV} & 22 {\rm PeV} & 162 {\rm PeV} \\
$ 10^{-1} \dot{M}$ & 11 {\rm GeV} & 2 {\rm PeV} & 16 {\rm PeV} \\
$ 10 \, \epsilon_B$ & 80 {\rm GeV} & 22 {\rm PeV} & 162 {\rm PeV} \\
$ 10^{-1} \epsilon_B$ & 11 {\rm GeV} & 2 {\rm PeV} & 16 {\rm PeV} \\
$ 10 \, n_A$ & - & - & - \\
$ 10^{-1} n_A$ & - & - & - \\
$ 10 \, T_{\rm age}$ & - & - & - \\
$ 10^{-1} T_{\rm age}$ & - & - & - \\
$2 \, \theta_{0}$ & 9 {\rm GeV} & 3 {\rm PeV} & 11 {\rm PeV} \\
$ 1/2 \,  \theta_{0}$ & 100 {\rm GeV} & 13 {\rm PeV} & 200 {\rm PeV} \\
$10^{-1} l_c$ & 266 GeV  & 4 PeV  & 130 PeV \\
$\eta=0.01$ & -  & -  & -  \\
$ \Delta_{\rm down} = l_c $ & 10 {\rm GeV} & 1 {\rm PeV} & 32 {\rm PeV} \\
$ \Delta_{\rm down} = 3 \, l_c $ & 23 {\rm GeV} & 4 {\rm PeV} & 47 PeV \\
$\alpha=1$  & - & - & - \\
$\alpha=2$ & - & - & 47 {\rm PeV} \\
\hline
\end{tabular}
\label{tab: param scan}
\tablefoot{Impact on the maximum energy of different parametric assumptions. In the first row, for reference we report the maximum energy for the three prototypes presented in Table~\ref{Tab: prototypes}. The first column reports the parameter modified with respect to the prototype. The three columns in the table refer respectively to protostar (PS), microquasar (MQ), and Seyfert (SEY). Missing entries in the table are meant to emphasize a lack of differences with respect to the prototype.}
\end{table}
\begin{figure}[t]
    \centering
    \includegraphics[width=1\columnwidth]{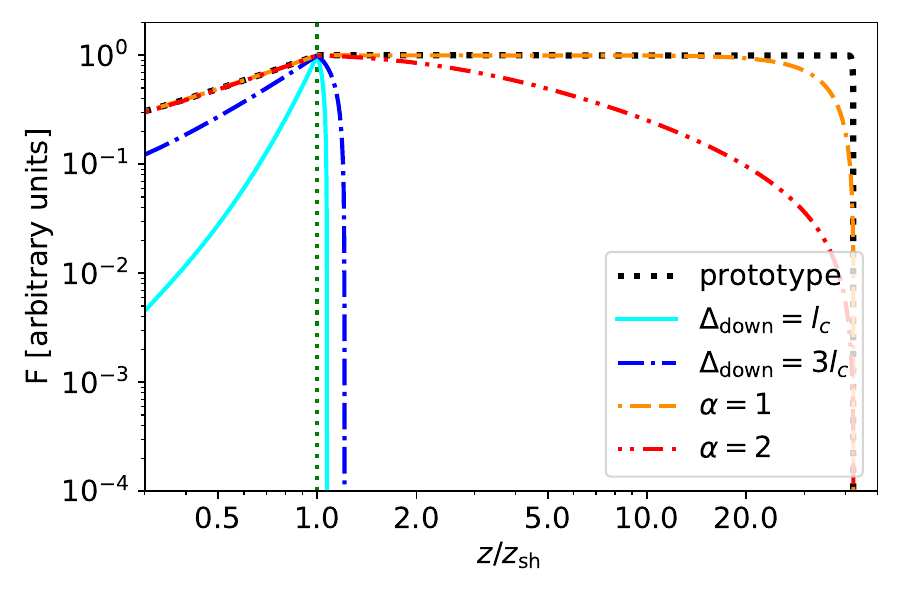} \quad \includegraphics[width=1\columnwidth]{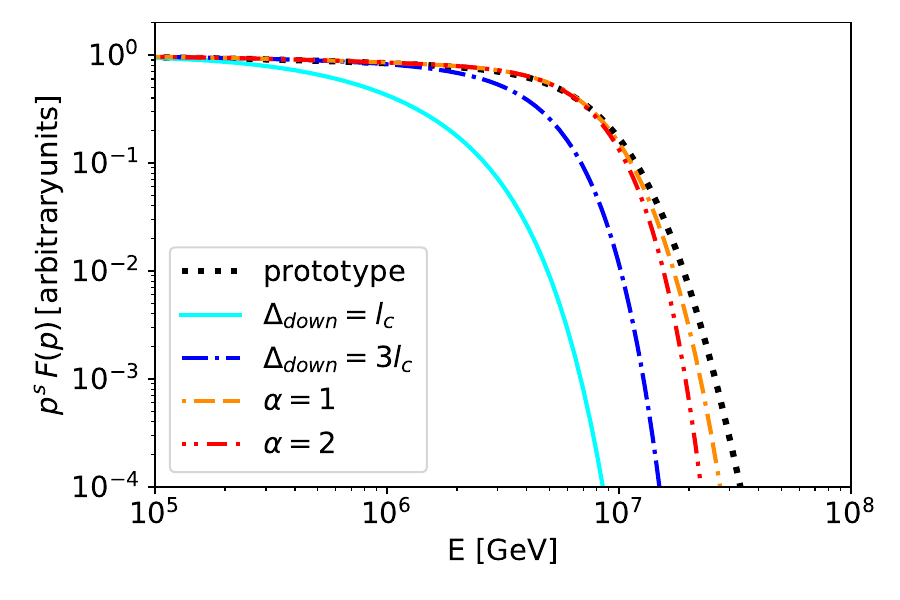}
    \caption{Impact of different downstream transport conditions for the case of MQ. \textit{Top}: Particle distribution along the jet. The vertical~dotted green line identifies the recollimation shock location. \textit{Bottom}: Spectrum of accelerated particles. The color-coding and line styles are identical for both plots. The solid cyan  line and the dot-dashed blue line represent  the exponential decay of the turbulence on typical scales $l_c$ and 3 $l_c$, respectively. The dash-dash-dotted orange line and double-dot-dashed red line represent respectively the power-law scaling diffusion coefficient with index 1 and 2. The dotted black line is the prototype MQ$_0$.}
    \label{Fig: Param variation MQ}
\end{figure}
Although selecting a prototype source offers a qualitative view of the range of maximum particle energies, variability within each jet class can span orders of magnitude—or at least significant factors — for the parameters considered. A complete exploration of the parameter space is beyond the scope of this work; instead, in this subsection we assess the impact of different parameter choices by varying each parameter individually relative to the SEY$_0$, MQ$_0$, and PS$_0$ prototypes, assuming Kraichnan diffusion.

In Table~\ref{tab: param scan} we report the results for $E_{\max}$ obtained by changing the parameters' values by some factor with respect to the benchmark cases. We first vary the velocity by a factor of 2 and the mass loss rate by a factor of 10.
In all scenarios we observe that the higher the kinetic power the higher the maximum energy. We approximately find that $E_{\rm max}$ scales with $\dot{M}^{1/2}$ in line with what we discussed in the previous subsection, whereas the scaling with velocity is not the same. In particular, for PS and SEY we observe a Kraichnan-like dependence, i.e., $\propto u_0^{2.5}$, while for the MQ the scaling is $\propto u_0^{1.5}$.
Such a different scaling derives from the maximum energy being in the transition region where $r_L \approx l_c$. 
Modifications of $\epsilon_B$ have identical scaling as the mass-loss rate. 
This is due to the assumption that the magnetic pressure is proportional to the ram pressure, which depends linearly on $\dot{M}$. 
Modifications of the density and age of the system do not affect $E_{\max}$ because the maximum energy does not depend significantly on the position of the recollimation shock, as Eq.~\eqref{eq: Hillas} shows for the case of Bohm diffusion. Also, this is a consequence of the assumption that $B^2 \propto \rho_0 u_0^2$ and on the additional assumption that $l_c$ is of the same order of the jet diameter.

Similarly to the case of the velocity, we observed that for different assumptions in $\theta_0$ we obtain a Bohm-like scaling for MQ, while SEY and PS feature a stronger dependence resulting from the Kraichnan-like scaling. 
We also explored the possibility of having a smaller coherence length by assuming it an order of magnitude smaller than the jet diameter. We observed that in the case of protostar and Seyfert this leads to an increase of $E_{\rm max}$, whereas in the case of microquasar the same operation results only in a small decrease. The former case is somewhat trivial, in that smaller $l_c$ results in smaller diffusion coefficient and consequently higher maximum energy. In the latter case, the decrease in the diffusion coefficient normalization is compensated by the transition to the small pitch angle scattering occurring at lower energies.

Changing $\eta$ to values $<1$ increases the distance of the recollimation shock (but this does not affect $E_{\max}$ as already discussed above) and decreases the height of the jet head, thereby reducing the size of the downstream. 
The latter effect may reduce the maximum energy only if the effective downstream size becomes significantly smaller than the one upstream, as discussed in the following. 

We explored different assumptions on the downstream turbulence. We consider two possible scenarios, each of them with two different scalings. 
The first one assumes that the turbulence decays exponentially throughout the downstream on a typical length-scale comparable to $l_c$. We implement it by moving the escaping boundary from the jet head to a distance, $\Delta_{\rm down}=l_c$ and 3 $l_c$ from the shock surface.   
The second scenario assumes a power-law decay of the turbulence in the downstream region resulting in a diffusion coefficient $D\sim z^\alpha$ and we explore it by considering a linear and quadratic scalings, i.e $\alpha=1$ and 2.
In Fig.~\ref{Fig: Param variation MQ} we report the spatial distribution of particles (top panel) and the spectra at the recollimation shock (bottom panel) associated with the different transport scenarios considered for the downstream region.
For the cases of exponentially decaying turbulence we observe a sizable decrease of the maximum energy, indicating that propagation in the downstream becomes the dominant effect in determining the maximum energy of the system. 
For the case of power-law decay we observe only a marginal difference in the shape of the cutoff, but the effective maximum energy remains substantially unchanged, meaning that the propagation upstream provides the most stringent constraint. 
In the case of $D_1\propto z^2$ we observe a mild decrease of $E_{\rm max}$ only in the case of Seyfert jets suggesting that, depending on the jet class, the conditions of confinement might become comparable between upstream and downstream.

\subsection{Comments on the particle escape from the system and hadronic gamma-ray production regions}
\label{subs: comments}

The particle transport is in general dominated by the advection with a subdominant impact of diffusion. 
For all jet classes of our interest, 
energy losses due to proton-proton collisions are negligible within the jet.
Therefore, one can expect that the majority of accelerated protons will either escape the system by crossing the jet head and the forward shock or be advected sideways in the cocoon after crossing the reverse shock and, possibly, the jet lateral surface. 

The ultimate fate of high-energy particles in the context of our model depends on the physical and micro-physical conditions they encounter near the outer boundary of the jet, namely at the jet head.
A detailed characterization of the jet-head structure, its microphysics, and the lateral advective flows feeding the cocoon goes beyond the scope of this work. However, they can naturally be expected to play a central role---together with a possible lateral escape from the jet---in the way relativistic particles populate the cocoon. 

The external cocoon, in turn, is expected to contain enough material to provide an effective target for pp interactions. 
However, the possibility for such a cocoon to shine in gamma rays highly depends on the local diffusion coefficient \citep[see also][for similar discussion in the context of microquasar remnants]{Abaroa26_MQR} and the possible presence of large scale flows connected to the cocoon intrinsic dynamics, an accretion disk wind or the central engine itself. 

All these aspects connected to the multiwavelength and high-energy neutrino emission will be explored in a forthcoming publication. 
However, it is reasonable to expect that the morphology of hadronic (pp) emission from a jet-cocoon system shall appear as a wide region extending along the whole cocoon size.
A somewhat more exotic hadronic channel may come from $p\gamma$ interactions in the vicinity of the central engine. In fact, a nonnegligible amount of protons in the cocoon, could be either advected in the backflows or diffuse near the jet launching region where photon fields are strong enough to provide a sizable target. $p\gamma$ interactions might also take place efficiently if the size of the system is extremely compressed, possibly due to a combination of young age and high density of the external medium.
On the other hand, depending on the local magnetic field encountered, the leptonic emission, may appear mainly in bright knots and/or localized along the jet spine similarly to what is observed in the case of SS433 \citep{HESS_SS433_2024,Tsuji_2025}.

{Finally, it is worth highlighting that the larger the inclination of the recollimation shock cusp, the smaller the available energy dissipated at the shock that can be converted into cosmic rays. Therefore, one can expect that the most luminous systems are those featuring a Mach disk.}

\section{Discussion and future prospects}
\label{Sec: discussion}

The treatment of particle acceleration and transport performed in this work aims at describing the leading-order physics of diffusive shock acceleration at recollimation shocks within a tractable semi-analytical framework. This choice, although it  allows  a systematic exploration of parameter space, naturally entails a number of simplifications that we discuss below.

First, the transport equation is solved in a 1D geometry along the jet axis. This approximation neglects transverse gradients and dimensional effects, including lateral particle escape.
In realistic jets, especially in nonmagnetically dominated flows, perpendicular transport and shear-induced effects may contribute to shaping both the particle spectrum, maximum energy and the spatial distribution of nonthermal emission. Therefore, including multidimensional transport and anisotropic diffusion coefficients will be an important step forward.
The lateral escape, in particular, might play a substantial role if the Larmor radius approaches the jet diameter---assumed also to be equal to the coherence length. In this context, the existence of regular magnetic field---especially a toroidal component---could strongly limit this effect. We believe that this aspect can motivate further investigations and dedicated simulations. 

Second, the present model does not explicitly include a dynamically relevant ordered guiding magnetic field, nor does it account for a strong toroidal component. While this is consistent with the assumption of a kinetically dominated jet at the scales of interest, different magnetic configurations could alter, as said, both the confinement properties and the acceleration efficiency. Extending the framework to include prescribed magnetic geometries and anisotropic diffusion represents a natural improvement.

Related to this point, the obliquity of the recollimation shock is treated only approximately by finding an effective angle of impact between the fluid and the shock. 
The complexity of particle acceleration at oblique shocks, and its potential implications on the spectral slope and acceleration efficiency when also guide magnetic field are included \citep{Kirk1989,Gieseler1999,Bell2011,Kirk2023,Reville2025} cannot be fully captured within the current formulation. 
In particular, the interplay between magnetic field orientation, shock geometry, and particle scattering may introduce potential deviation from the canonical spectral indices discussed here. A more self-consistent treatment of shock obliquity, its extension and magnetic field topology, possibly informed by kinetic or hybrid simulations, will be the focus of future investigations.

The model focuses on acceleration at the recollimation shock, while strong shocks may also be present at the jet head, where both reverse and forward shocks can contribute to nonthermal particle production, or possibly lead to a re-acceleration scenario. Although this does not invalidate the present scenario, a global treatment coupling recollimation and jet head would provide a more comprehensive picture of particle energization in collimated outflows. 
This aspect will be the focus of a follow-up work in preparation.
{In addition, as shown in numerical MHD simulations\citep{Bodo2018,Sciaccaluga2025,Boula2025,Costa2026}, the jet geometry after the first recollimation shock can develop strong instabilities and strong reflection shocks, where particles can also be accelerated. In this context, the jet could feature a train of recollimation shocks, typically accompanied by secondary reflected shocks depending on the pressure balance with the surrounding medium.}

Overall, these limitations do not affect the internal consistency of the model, but rather delineate a well-defined domain of applicability and highlight clear directions for systematic extensions, including multidimensional transport in different magnetic configurations, oblique shock physics, and relativistic generalizations.

The main limitation of the hydrodynamical description is the assumption of stationarity. Both the jet central engine and the external medium are treated as steady, and the ambient density is assumed to be uniform. While this represents a reasonable first-order approximation and can often be mapped onto effective time-averaged quantities, it cannot capture the intrinsic variability that could characterize some AGN, microquasars, and protostellar systems. 
Time-dependent activity of the central engine, intermittency, and stratified or magnetized external media (e.g., wind-like density profiles) may significantly affect the location and strength of recollimation shocks as well as producing interesting 3D effects such as kink and sausage instability where other internal shocks could develop. A time-dependent MHD extension of the present framework in complex media would therefore be highly valuable.

The model is developed in the subrelativistic regime, as an extension of the \cite{Bromberg_2011} formalism. As a consequence, it can be safely applied only up to mildly relativistic velocities. This is particularly relevant for AGN jets and microquasars. Many objects in these source classes can reach, in fact, relativistic bulk speeds. A fully relativistic generalization would allow the framework to address more powerful systems, including relativistic microquasars, Fanaroff–Riley (FR) AGN jets and blazars, which are characterized by high kinetic luminosities and relativistic launching speeds. 
Interestingly, observations indicate that FR jets, while initially relativistic, often show strong deceleration at kiloparsec scales \citep{Kataoka_2008,Mullin_2009,Reddy_2023}, precisely where strong recollimation shocks are expected to develop \citep{Laing_2002}. Hydrodynamical simulations \citep[e.g.,][]{Perrucho_2007,Perucho_2014} suggest that such shocks can be dynamically important and, in some cases, even disrupt the flow \citep[see also][]{Perrucho_2019_,Perrucho2022}. Extending the present semi-analytical approach to this relativistic regime represents a promising avenue for future work.

Additional source-dependent limitations should also be noted. In protostellar jets, high densities may imply that the condition $\tilde{L} \ll 1$ is not always satisfied, even at large distances from the source, potentially modifying the recollimation properties. Moreover, magnetic fields may play a nonnegligible role in the collimation process itself \citep{Ustamujic_2016}, requiring a magnetohydrodynamical treatment.

In microquasars associated with high-mass X-ray binaries, the interplay between jet and stellar wind can further complicate the dynamics. If the wind luminosity exceeds the cocoon pressure, the global shock structure and the formation of recollimation shocks may differ from the simplified scenario considered here, such as in the case of Cygnus X-3, one of the most powerful Galactic PeV accelerators \citep{LHAASO_CygnusX-3}, where the massive companion is a Wolf-Rayet star \citep{Koljonen2017}.

Finally, it is worth noting that geometrical configurations similar to those discussed in this work, are not exclusive of recollimation shocks, but can generically occur in internal shocks in jets.

\section{Conclusions}
\label{Sec: Conclusions}

In this work we investigated particle acceleration via diffusive shock acceleration at recollimation shocks in subrelativistic jets. 
In order to do so, we formulated a subrelativistic extension of the hydrodynamics model of the jet-cocoon system as developed by \cite{Bromberg_2011}.
In such a framework, we studied acceleration and transport by solving the space-dependent transport equation in the jet assuming global stationarity and free escape at the jet head and zero net flux at the launching point of the jet. 
Our model is versatile and was applied to different source classes: Seyfert jets, microquasars, and protostellar jets. 

We explored these different jet classes by setting a prototype for each of these objects, and we investigated the impact of different assumptions for particle diffusion. 
We focused our attention on the spectral shapes and maximum energy of the accelerated particles. 
We found simple analytic expressions for upper limits to such a maximum energy in each source class, finding that Seyfert can accelerate from PeV up to EeV, microquasars up to tens of PeV, and protostellar jets up to the TeV range. 

We explored the parameter space for each source class, finding results in good agreement with our analytic estimates. 
In general, we found that the upstream region of the recollimation shock poses the most stringent conditions on the maximum available energies.
In this context, we also explored the impact of spatially decaying turbulence in the downstream region of the recollimation shock concluding that in certain configurations, the limits imposed by the downstream region might become the dominant ones.

We conclude with an observational perspective. 
Recollimation shocks and, more generally, internal shocks in collimated jets, might appear as bright knots and shine because of leptonic emission. 
In this work, we clearly showed that proton inelastic collisions are not expected to be relevant in subrelativistic collimated jets. 
However, protons might leak from the jet through either the jet head or laterally and diffuse in the denser cocoon.
Depending on the local environmental conditions encountered in cocoons, inelastic proton-proton, and/or possibly $p\gamma$, collisions might result in extended morphologies of hadronic nature. 
{Finally, recollimation shocks with Mach disks are likely more efficient accelerators than those featuring highly inclined cusps.}

\begin{acknowledgements}
       EP acknowledges economic support from INAF through “Assegni di ricerca per progetti di ricerca relativi a CTA e precursori”. Part of this work was carried out at the APC Laboratory, whose hospitality EP gratefully acknowledges. EP is grateful to M. Pais, A. Zegarelli, S. Celli and M. Lemoine for insightful discussions on jet physics. E.P. is also grateful to L. Foschini and F. Tavecchio for valuable comments helping to improve the manuscript. GM and EA are partially supported by the INAF Theory Grant 2024 {\it Star Clusters As Cosmic Ray Factories II}. SSC acknowledges support from the ANR grant ``MiCRO'' with the reference number ANR-23-CE31-0016.
\end{acknowledgements}

\bibliographystyle{aa} 
\bibliography{bibliography} 

\appendix

\section{Similarities between jets and wind bubbles}
\label{Subs: Appendix Bubble similarity}

It is worth noticing that the jet-cocoon system has the same temporal evolution as an adiabatic wind-blown bubble \cite[]{Weaver_1977}. More specifically, the recollimation shock is analogous to the wind termination shock, and the jet head and the cocoon expansion evolve like the forward shock in the wind bubble. Such an analogy is to be expected, as both systems are powered by a central source with constant kinetic luminosity. 
We also note that the total energy of the jet, $L_j t$, and the one of the swept-up mass, $M_{c} v_c^2$ have identical scalings with time. 
Therefore, for the whole period of (steady) activity of the central engine, the cocoon expansion is expected to remain self-similar, regardless of the total mass in the cocoon. 

The same scalings obtained in Sect.~\ref{Sec: dyn-and-structure} arise from solving the momentum and energy conservation equations for the shells, at the jet head ($\dot{M}_{c,h}$) and on the side ($\dot{M}_{c,s}$), driven by the cocoon pressure under self-similar conditions:
\begin{align}
    &\partial_t[{M}_{c,h} v_h] = \pi R_c^2 P_c \chi \\
    &\partial_t[{M}_{c,s} v_c] =  2 \pi R_c z_h  P_c \\
    &\partial_t{E}_{c} = L_j - \left[ \pi R_c^2 v_h \frac{r_{RS}}{(r_{RS}-1)} + 2 \pi R_c z_h v_c \right] P_c, 
\end{align}
where $\chi=\rho_A v_h/P_c$ is the ratio among the ram pressure working on the jet head and the cocoon pressure. $R_c$ and $z_h$ are the cocoon radius and jet head location respectively.
While this approach provides a good estimate of the system scalings ($z_h \sim R_c \sim t^{3/5}$ and $P_c \sim t^{-4/5}$), it does not provide the location of the recollimation shock.

\section{Sink term in the transport equation}
\label{Subs: Appendix sink term}

In principle, using the formalism presented in this work, it is possible to account also for energy losses or particle escape along the lateral jet surface by introducing a sink term of the form $f/\tau$
on the right-hand side of Eq.~\eqref{Eq. Transport}.

The inclusion of such a term allows us to write the final solution for the distribution function in the same form with the following substitutions:
\begin{align}
    &G_0(z,p) \longrightarrow G_0(z,p) + H_0(z,p) \\
    &G_1(z,p) \longrightarrow G_1(z,p) + H_1(z,p),
\end{align}
where
\begin{align}
    &H_0(z,p) = \int_0^z dz' f_0(z',p)/ \tau(z',p) \\
    &H_1(z,p) = \int_z^{z_{\rm esc}} dz' f_1(z',p)/ \tau(z',p).
\end{align}

\section{Impact of the turbulence cascade in the downstream region:  Analytic solutions}
\label{Subs: Appendix Turbulence cascade}

Given the geometry of the jet, the most natural approximation for the downstream region is the cylindrical one.
In this context, the area $A(z)$ becomes a constant.
For the flux conservation equation   the downstream speed $u_1(z)$ will also remain constant for the whole downstream region. 

Now, if the downstream turbulence decays on scales much larger than the jet size, the downstream diffusion coefficient, $D_1$ can be considered constant and solution can be expressed in the following analytic form:
\begin{align}
\label{Eq. down-const}
    &f_{1}(z,p) = f_{\rm sh}(p) \frac{1-e^{\mathcal{Y}(z,p)-\mathcal{Y}(z_{\rm esc},p)}}{1-e^{-\mathcal{Y}(z_{\rm esc},p)} } \\
    &\mathcal{Y} = \frac{u_1}{D_1(p,z)} (z-z_{\rm sh}).
\end{align}
Under the uniform cylindrical approximation, one can show that also the downstream high-energy cutoff function assumes the following simplified form:
\begin{equation}
\label{Eq. down-cut-const}
    \Gamma_{1}(p) = \frac{s}{r} \int_{p_{\rm inj}}^p \frac{dp'}{p'} \frac{1}{\exp[\mathcal{Y}(z_{\rm esc},p)]-1}
.\end{equation}

However, the magnetic turbulence may decay on spatial scales shorter than the jet size and it can be useful to provide the full expressions for such a case. As an example we consider here two possible scenarios: linear decay, i.e $D_1(p,z) = D_0(p) (z/z_{\rm sh})$, and quadratic decay, $D_1(p,z) = D_0(p) (z/z_{\rm sh})^2$.
For the linear decay, the downstream solution and the cutoff function result in the following expressions:
\begin{align}
\label{Eq. down-lin}
    &f_{1,\rm lin}(z,p) = f_{\rm sh}(p) \frac{1-(z/z_{\rm esc})^{u_1z_{\rm sh}/D_0(p)}}{1-(z_{\rm sh}/z_{\rm esc})^{u_1z_{\rm sh}/D_0(p)}} \\
    &\Gamma_{1,\rm lin}(p) = \frac{s}{r} \int_{p_{\rm inj}}^p \frac{dp'}{p'} \frac{1}{(z_{\rm esc}/z_{\rm sh})^{u_1z_{\rm sh}/D_0(p)}-1},
\label{Eq. cut-down-lin}
\end{align}
where $D_0(p)$ is the diffusion coefficient in the immediate downstream of the recollimation shock.
We note that the functional forms of the solution for $D\sim z$ and $D\sim z^0$ are identical;   the only exception is the scaling function $\mathcal{Y}$, which is replaced by 
\begin{equation}
    \mathcal{Y}_{\rm lin} = \frac{u_1 z_{\rm sh}}{D_0(p)} \ln{\left(\frac{z}{z_{\rm sh}}\right)}.
\end{equation}
Finally, it is also possible to show that the quadratic decay scenario has identical functional form to the case of constant $D_1$, namely Eqs. \eqref{Eq. down-const} and \eqref{Eq. down-cut-const}, but with a different scaling function $\mathcal{Y}_{\rm quad}$ that is modified as in the following expression:
\begin{equation}
    \mathcal{Y}_{\rm quad} = \frac{u_1 z_{\rm sh}}{D_0(p)} \left(1-\frac{z_{\rm sh}}{z}\right).
\end{equation}

\section{Semi-analytic iterative algorithm for the solution of the transport equation}
\label{Subs: Appendix algorithm}

As discussed in Appendix~\ref{Subs: Appendix Turbulence cascade}, the downstream solution, $f_1(z,p)$, depends only on the spectrum at the shock, $f_{\rm sh}(p)$, and on analytic functions. However, neither the upstream solution nor the shock spectrum admits a simple analytic expression, since they depend in a nonlinear way on each other according to Eqs.~\eqref{Eq. Up-Sol} and~\eqref{Eq. Shock solution}, respectively. 
In particular, the nonlinearity is embedded in the term $u_{\rm eff,0}$ for the former, and in $\Gamma_0$ for the latter.

In order to find the solutions to Eqs.~\eqref{Eq. Up-Sol} and \eqref{Eq. Shock solution}, we employ a recursive algorithm. 
We define as guess functions the expressions characterized by a null implicit part as follows:%
\begin{align}
    &f_{\rm sh}^{(0)}(p) = \eta_{\rm eff} \frac{s n_0}{4 \pi p_{\rm inj}^3} \left( \frac{p_{\rm inj}}{p} \right)^s e^{-\Gamma_0(p)}, \\
    &f_0^{(0)}(z,p) = f_{\rm sh}(p) \exp{\left[ -\int_z^{z_{\rm sh}} dz' \frac{u_{\rm 0}(z')}{D_0(z',p)} \right]}.
\end{align}
From the latter expression, at the first step we can compute $g_0$, $G_0$, $u_{\rm eff, 0}$ and $\Gamma_0$ from which new expression of $f_{\rm sh}$ and $f_0$ can be obtained.
At the generic step $j+1$ the solution at the shock and in the upstream read
\begin{align}
    &f_{\rm sh}^{(j+1)}(p) = \eta_{\rm eff} \frac{s n_0}{4 \pi p_{\rm inj}^3} \left( \frac{p_{\rm inj}}{p} \right)^s e^{-\Gamma_0(p)} e^{-\Gamma_1^{(j)}(p)}, \\
    &f_0^{(j+1)}(z,p) = f_{\rm sh}(p) \exp{\left[ -\int_z^{z_{\rm sh}} dz' \frac{u_{\rm eff,0}^{(j)}(z')}{D_0(z',p)} \right]}.
\end{align}
By comparing $f^{(j+1)}$ with $f^{(j)}$ one can check the convergence, which is typically achieved in a few tens of iterations. 
At that point, when the convergence is reached, we compute $f_1(z,p)$ following the prescriptions described in Appendix~\ref{Subs: Appendix Turbulence cascade}.

\section{Analytic considerations}
\label{Appendix: analytic considerations}

For illustrative purposes let us analyze the qualitative properties of the upstream solution assuming a divergence free wind in the jet, namely an idealistic configuration in which $\partial_z[A u] = 0$. 
For simplicity we also assume a space-independence of the diffusion coefficient.   
The immediate result of such an assumption is that the functions $G_i$ goes to zero. This reduces the upstream solution, Eq.~\eqref{Eq. Up-Sol}, to the well-known result of the infinite planar shock:
\begin{equation}
    f^*_{0}(z,p) = f_{\rm sh}(p) \exp{\left[- (z_{\rm sh}-z) u_0/D_0(p) \right ]}.
\end{equation}

A further consistency check of our solution can be obtained by considering the full spherical expansion, namely when $A(z) \sim z^2$. For the sake of simplicity, let us restrict our attention to the case of adiabatic shocked wind, such that $u_1(z) \sim z^{-2}$.
It is straightforward to see that the wind in the region is divergence-free, so that $G_1 = 0$ and $u_{\rm eff,1}(z,p) = u_1(z)$.
With a bit of algebra it is possible to show that the function $\mathcal{I}(z,p)$ assumes the following analytic expression:
\begin{align}
    &\mathcal{I}^*(z,p) = \frac{1}{u_1 z_{\rm sh}^2} \left\{ 1 - \exp{\left[ -\mathcal{Y^*}(z,p) \right]} \right\} \\
    &\mathcal{Y^*}(z,p) = \frac{u_1 z_{\rm sh}}{D_1(p)} \left( 1 - \frac{z_{\rm sh}}{z}\right) .
\end{align}
Substituting the latter expression in the downstream solution one obtains the following result:
\begin{equation}
    f_{1,\rm sph}^*(z,p) = f_{\rm sh}(p) \frac{1-e^{\mathcal{Y^*}(z,p)-\mathcal{Y^*}(z_{\rm esc},p)}}{1-e^{-\mathcal{Y^*}(z_{\rm esc},p)} }
.\end{equation}
This coincides with the solution provided in \cite{Morlino2021} for the case of the adiabatic stellar cluster wind.

\end{document}